\documentclass[
	       twocolumn,
               prd,nofootinbib,superscriptaddress,
               showpacs,preprintnumbers,
               amsmath,amssymb]
              {revtex4}
\usepackage{graphicx,color}

\newcommand{\Tr}{\mathop\mathrm{Tr}\nolimits}
\newcommand{\Real}{\mathop\mathrm{Re}\nolimits}
\newcommand{\Imag}{\mathop\mathrm{Im}\nolimits}
\newcommand{\Dm}{D_\mu}
\newcommand{\Dmu}{D^\mu}
\newcommand{\dm}{\partial_\mu}
\newcommand{\dn}{\partial_\nu}
\newcommand{\cB}{\mathcal{B}}

\newcommand{\bra}[1]{\langle #1 |}
\newcommand{\ket}[1]{| #1 \rangle}
\newcommand{\bm}[1]{\boldsymbol{#1}}
\newcommand{\bk}{{\bm{k}}}
\newcommand{\bkind}{^{\vphantom{*}}_\bk}

\newcommand{\fk}{f^{\vphantom{*}}_\bk}
\newcommand{\fmk}{f^{\vphantom{*}}_{-\bk}}
\newcommand{\gsk}{g^{*}_\bk}
\newcommand{\half}{\frac{1}{2}}
\newcommand{\e}{\mathrm{e}}
\newcommand{\const}{\mathrm{const}}

\newcommand{\vpar}[2]{\frac{\delta #1}{\delta #2}}

\newcommand{\TeV}{\mathrm{ TeV}}
\let\eq=\eqref
\providecommand{\href}[2]{#2}

\begin{document}

\preprint{INR/TH-2003-5, BUHEP-03-09}

\title{Semiclassical Study of Baryon and Lepton Number Violation in
  High-Energy Electroweak Collisions.}

\author{F.~Bezrukov}
\email{fedor@ms2.inr.ac.ru}
\affiliation{Institute for Nuclear Research of the Russian Academy of Sciences,\\
  60th October Anniversary prospect 7a, Moscow 117312, Russia}
\author{D.~Levkov}
\email{levkov@ms2.inr.ac.ru}
\affiliation{Institute for Nuclear Research of the Russian Academy of Sciences,\\
  60th October Anniversary prospect 7a, Moscow 117312, Russia}
\affiliation{Moscow State University, Department of Physics,\\
  Vorobjevy Gory, Moscow, 119899, Russian Federation}
\author{C.~Rebbi}
\email{rebbi@bu.edu}
\affiliation{Department of Physics---Boston University\\
  590 Commonwealth Avenue, Boston MA 02215, USA}
\author{V.~Rubakov}
\email{rubakov@ms2.inr.ac.ru}
\affiliation{Institute for Nuclear Research of the Russian Academy of Sciences,\\
  60th October Anniversary prospect 7a, Moscow 117312, Russia}
\author{P.~Tinyakov}
\email{Peter.Tinyakov@cern.ch}
\affiliation{Institute of Theoretical Physics, University of Lausanne,\\
  CH-1015 Lausanne, Switzerland}
\affiliation{Institute for Nuclear Research of the Russian Academy of Sciences,\\
  60th October Anniversary prospect 7a, Moscow 117312, Russia}

\date{April 18, 2003}

\begin{abstract}
  We make use of a semiclassical method for calculating the
  suppression exponent for topology changing transitions in
  high-energy electroweak collisions.  In the Standard Model these
  processes are accompanied by violation of baryon and lepton number.
  By using a suitable computational technique we obtain results
  for s-wave scattering
  in a
  large region of initial data.  Our results show that baryon and
  lepton number violation remains exponentially suppressed up to very
  high energies of at least 30 sphaleron masses (250~TeV).  We also
  conclude that the known analytic approaches inferred from low energy
  expansion provide reasonably good approximations up to the sphaleron
  energy (8~TeV) only.
\end{abstract}


\maketitle

\section{Introduction}
\label{sec:intro}

Non-perturbative phenomena related to tunneling are often encountered
in quantum field theory.  Well known examples are false vacuum decay
and instanton-like transitions (the latter are accompanied by
non-\hspace{0pt}conservation of fermion quantum numbers).  When these
phenomena are governed by a small coupling constant they can generally
be studied by semiclassical methods.  This is certainly the case at
low energy or in situations which involve large number of quanta in
the initial state.  At low energy, collision processes can be well
described by a semiclassical approximation relying on the existence of
classical Euclidean time solutions to the equations of motion
interpolating between initial and final states.  In the examples
mentioned above these are bounce~\cite{Coleman:1977py} and
instanton~\cite{Belavin:1975fg}, respectively.  The probability of the
process is then proportional to the exponent of the Euclidean action
of these solutions.  As the action is inversely proportional to the
(small) coupling constant, the processes are highly suppressed.  The
effect of low energy excitations in the initial state (colliding
particles) gives only a pre-exponential factor and is inessential.

The situation changes at high energy, namely, at energy of the order
of the tunneling barrier height which separates initial and final
states.  In general, there exists a static unstable solution to the
equations of motion that lies on top of the potential
barrier~\cite{Klinkhamer:1984di} (properly speaking, at a saddle point
of the potential).  In field theory this solution is often referred to
as the ``sphaleron'', a name which we will use throughout this paper.
The minimum height of the barrier is precisely the sphaleron energy
$E_\mathrm{sph}$.  Naively, from the analogy with quantum mechanics of one
degree of freedom, one would expect that at energy higher than the
sphaleron energy the exponential suppression disappears.  This is
indeed what happens at finite temperature~%
\cite{Kuzmin:1985mm,Arnold:1987mh,Arnold:1988zg,Bochkarev:1987wg,
Khlebnikov:1988sr,Grigorev:1989je,Kuznetsov:1997sf,Frost:1999eh,Bonini:2000mb},
finite fermion density~%
\cite{Rubakov:1985am,Rubakov:1986nk,Matveev:1986bw,Matveev:1987gq,
Diakonov:1992mp}, or in the presence of heavy fermions in the initial
state~\cite{Rubakov:1985ix,Ambjorn:1985bb,Rubakov:1985it}.  But in
high energy particle collisions this is not necessarily the case, due
to the fact that the characteristic size of the sphaleron
configuration is much larger than the wavelength of the incoming
particles.  At the same time the application of a semiclassical
technique becomes problematic because the initial state no
longer involves a large number of quanta.

As was first noted in Refs.~\cite{Ringwald:1990ee,Espinosa:1990qn}, at
relatively low energy the corrections to the collision-induced
tunneling rate can be calculated by perturbative expansion in the
background of the instanton.  Further studies showed that the
actual expansion parameter in most models, including electroweak
theory, is $E/E_\mathrm{sph}$~%
\cite{McLerran:1990ab,Khlebnikov:1991ue,Yaffe:1990iy,Arnold:1990va}
and the total cross section of the induced tunneling has an
exponential form
\begin{displaymath}
  \sigma_{tot} (E) \sim
  \exp \left\{
    -\frac{16\pi^2}{g^2} F_{HG}(E/E_\mathrm{sph})
  \right\} \;,
\end{displaymath}
where $g$ is the small coupling constant and the function\footnote{The
  subscript $HG$ here stands for ``holy grail'' \cite{Mattis:1992bj}.}
$F_{HG}(E/E_\mathrm{sph})$ is a series in fractional powers of
$E/E_\mathrm{sph}$ (for a review
see~\cite{Mattis:1992bj,Tinyakov:1993dr,Rubakov:1996vz}).

While the perturbation theory in $E/E_\mathrm{sph}$ is limited to
small $E$, the general exponential form of the total cross section
implies that there might exist a semiclassical-type procedure which
would allow, at least in principle, to calculate
$F_{HG}(E/E_\mathrm{sph})$ at $E\gtrsim E_\mathrm{sph}$.  However, since the
initial state of two highly energetic particles is not semiclassical,
the standard semiclassical procedure does not apply and a suitable
generalization is needed, which was proposed in
Refs.~\cite{Rubakov:1992fb,Tinyakov:1992fn,Rubakov:1992ec}
and further developed in Refs.~\cite{Kuznetsov:1997az,Bezrukov:2001dg}.
The corresponding formalism reduces the calculation of the exponential
suppression factor to a certain classical boundary value problem,
whose analytical solution is not usually possible.

This approach is based on the conjecture that, with exponential
accuracy, the two-particle initial state can be substituted by a
multiparticle one provided that the number of particles is not
parametrically large (although not proven rigorously, this conjecture
was checked in several orders of perturbation theory in
$E/E_\mathrm{sph}$ in gauge
theory~\cite{Tinyakov:1992fn,Mueller:1993sc} and explicitly in quantum
mechanics with two degrees of
freedom~\cite{Bonini:1999cn,Bonini:1999kj}).  The few-particle initial
state, in turn, can be considered as a limiting case of a truly
multiparticle one with the number of particles
$N=\tilde{N}\cdot(4\pi/g^2)$, when the parameter $\tilde{N}$ is sent
to zero.  For the multiparticle initial state the transition rate is
explicitly semiclassical and has the form
\begin{equation}
  \sigma(E,N) \sim \exp \left\{
  -\frac{16\pi^2}{g^2} F(E/E_\mathrm{sph},\tilde{N}) \right\} .
\end{equation}
According to the above conjecture, the function
$F_{HG}(E/E_\mathrm{sph})$, corresponding to the two-particle incoming
state, is reproduced in the limit $\tilde{N}\to 0$,
\[
  \lim\limits_{\tilde{N}\to 0}F(E/E_\mathrm{sph},\tilde{N})
  = F_{HG}(E/E_\mathrm{sph}).
\]
Therefore, albeit indirectly, the function
$F_{HG}(E/E_\mathrm{sph})$ is also calculable semiclassically.

Within the semiclassical framework, the function
$F(E/E_\mathrm{sph},\tilde{N})$ is determined by the action evaluated
at a particular solution to the classical field
equations~\cite{Rubakov:1992ec} on a certain contour in complex time.
In this formulation, the problem, at least in principle, is amenable to
a computational solution.  Namely, one has to solve the corresponding classical
boundary value problem numerically and calculate the function
$F(E/E_\mathrm{sph},\tilde{N})$, which then can be used to extract
information about $F_{HG}(E/E_\mathrm{sph})$.

The implementation of this technique is nevertheless highly
non-trivial.  The differential equations one encounters
are partially of the hyperbolic type (along the Minkowskian
parts of the time contour) and partially of elliptic type (along the
Euclidean part), which makes their numerical solution particularly
challenging.  In the electroweak theory, additional difficulties arise
from the need to deal with the large number of internal degrees of
freedom, unphysical modes due to gauge invariance, and time
translational symmetry which cause an unwelcome degeneracy in the
numerical procedure used to find the semiclassical solutions.

Moreover, at high energy (roughly, at energy higher than the sphaleron
energy) tunneling solutions, interpolating between vicinities of
different vacua in finite time, cease to exist.  This subtlety turned
out to be a general problem in the description of tunneling in systems
with many degrees of freedom, and it has to do with the nontrivial way
tunneling occurs at high energy---the system prefers to create a state
close to the sphaleron, which then decays into the correct vacuum.  To
find the corresponding suppression exponent numerically one has to use
a properly regularized version of the boundary value problem,
developed in Ref.~\cite{Bezrukov:2003yf}.

In a long program of investigations we have been able to gradually
overcome all of these hurdles.  Preliminary results for energies below
the sphaleron energy were reported in Ref.~\cite{Bezrukov:2001dg}.
Now we are in a position of presenting what we are confident is the
full solution to the numerical problem for a wide range of energy and
incoming particle number, including energies above the sphaleron.
The field configurations we analyse in this paper are restricted  to
be spherically symmetric in space.  Hence our results apply, strictly
speaking, to s-wave scattering only.

In this paper we will concentrate on obtaining the suppression
exponent for collision-induced tunneling in $SU(2)$ gauge model with
the Higgs mechanism, corresponding to the electroweak sector of
the Standard Model at zero $\theta_W$.  The problem is particularly
interesting because of the baryon number violation which accompanies
such processes \cite{'tHooft:1976fv} and the relatively low sphaleron
energy $E_\mathrm{sph}\simeq8\TeV$.  Though computational limitations
do not allow to reach literally zero value of the rescaled number of
particles $\tilde{N}$, corresponding to particle collisions, we were
able to extrapolate the results to zero $\tilde{N}$ and get a
bound on the suppression exponent
(strictly speaking, for s-wave scattering)
and also provide an
estimate for this exponent.

In section~\ref{sec:formulation} we present the detailed formulation
of the problem, outline the method and present the main physical
results.  In section~\ref{sec:rst} we give the derivation of the
semiclassical method for the gauge model.  The lattice formulation of
the equations and subtleties appearing in the discretized version are
given in section~\ref{sec:lattice}.  Application of the regularization
method of Ref.~\cite{Bezrukov:2003yf} is described in
section~\ref{sec:reg}.  Detailed numerical results are presented in
section~\ref{sec:num_results}.  Our conclusions are in
section~\ref{sec:conclusions}.

\section{Formulation of the problem and main results}
\label{sec:formulation}

Non-Abelian gauge models have an infinite number of topologically
distinct vacua, labeled by an integer topological number.  Processes
changing the topological number are accompanied by violation of
fermion (baryon and lepton) numbers \cite{'tHooft:1976fv}, a
phenomenon of great interest for cosmology and particle physics.  The
topologically distinct vacua are separated by a potential barrier,
whose height, in models with the Higgs mechanism, is given by the
sphaleron energy.  Topology changing transition may occur via
tunneling at low energies or, at sufficiently high energy and suitable
initial state, via classical evolution over the sphaleron.

In this paper we study a four-dimensional model which captures all the
important features of the Standard Model---an $SU(2)$ gauge theory
with the Higgs doublet.  This model corresponds to the bosonic sector
of the Standard Model with $\theta_W=0$.  To the leading order in the
coupling constant, the effect of fermions on the gauge and Higgs
fields dynamics can be ignored~\cite{Espinosa:1992vq}.  The action of
the model is
\begin{align}\label{SU2action}
  S = \frac{1}{4\pi\alpha_W}
      \int\! d^4x \left\{\vphantom{\frac{1}{2}}\right.
                         &
        -\frac{1}{2}\Tr F_{\mu\nu}F^{\mu\nu}
  \\
  & \left.\vphantom{\half}
      +(\Dm\Phi)^\dagger\Dmu\Phi
      -\lambda(\Phi^\dagger\Phi-1)^2
    \right\}
  \;, \nonumber
\end{align}
where
\begin{align}
  F_{\mu\nu} &= \dm A_\nu-\dn A_\mu-i[A_\mu,A_\nu] \\
  \Dm\Phi    &= (\dm-iA_\mu)\Phi
\end{align}
with $A_\mu=A_\mu^a\sigma^a/2$ and $\alpha_W=g^2/4\pi$.  Here we have
eliminated inessential constants by an appropriate choice of units.
The dimensional parameters can be restored noting that in the
normalization \eqref{SU2action}, the gauge boson mass is
\begin{equation}\label{mWchoice}
  M_W=\frac{1}{\sqrt{2}}\;,
\end{equation}
and the Higgs boson mass is
\begin{equation*}
  M_H=\sqrt{8\lambda}M_W \;.
\end{equation*}
In most of our calculations the Higgs self-coupling $\lambda$ was set
equal to $\lambda=0.125$, which corresponds to $M_H=M_W$.  The
dependence on the Higgs boson mass is very weak, so this is a
reasonable approximation.  Also we often omit the omnipresent overall
factor $1/\alpha_W$.

Any vacuum configuration in this model can be obtained from the
trivial vacuum $A_\mu=0$, $\Phi=\Phi_v=\begin{pmatrix}0 \\
1\end{pmatrix}$ by a certain gauge transformation $U(x)$.  We will be
using the temporal gauge $A_0=0$, where the vacuum configurations are
described by time-independent $U(\bm{x})$, corresponding to residual
gauge invariance.  In this gauge, field values at spatial infinity
cannot change during the evolution (otherwise the kinetic term becomes
infinite) and thus one considers only those $U(\bm{x})$ which have
some fixed asymptotics at spatial infinity.  Often the asymptotic
$U(\bm{x}\to\infty)\to1$ is used, so any vacuum configuration
corresponds to a mapping from space $R^3$ with identified infinity,
which is homotopically equivalent to $S^3$, to the gauge group
$SU(2)\sim S^3$.  The degree of this mapping is precisely the
topological number of the corresponding vacuum.  A gauge choice of
this form is convenient for analysis of the excitations about the
trivial vacuum.  For other purposes it may however be useful to choose
an alternative behavior of the gauge function at spatial infinity,
like $U(\bm{x})\to \exp\{i\boldsymbol{\bm{\sigma x/|x|}}\}$, which maps
the $S^2$ of spatial infinity to the equatorial $S^2$ of the $SU(2)$.
The two neighboring vacua then map the space $R^3$ either to north or
south hemisphere of the $SU(2)$.  In this gauge, the sphaleron
configuration takes the simplest form, and we will use this gauge
everywhere in this paper, except for the analysis of the mode
expansion in the initial state.

Numerous perturbative attempts were made to find the probability of
the collision-induced topology changing transitions in this model (see
Refs.~\cite{Mattis:1992bj,Tinyakov:1993dr,Rubakov:1996vz} for
reviews), giving reliable results only for relatively low energies.  A
non-perturbative study of classically allowed over-barrier transitions
was presented in Ref.~\cite{Rebbi:1996zx}.  All solutions found in
Ref.~\cite{Rebbi:1996zx} are configurations with large numbers of
particles in the initial state and thus they do not correspond to
realistic collisions.  Another approach, pursued in this paper, is to
use the semiclassical method of Refs.~%
\cite{Rubakov:1992fb,Tinyakov:1992fn,Rubakov:1992ec,Kuznetsov:1997az} adapted for
theories with gauge degrees of freedom.  This method was implemented
in Ref.~\cite{Bezrukov:2001dg}, where the results were obtained for
energies below $E_\mathrm{sph}$, what suggested that at the sphaleron
energy the suppression is still strong.  However, a straightforward
application of the technique of Refs.~%
\cite{Rubakov:1992fb,Tinyakov:1992fn,Rubakov:1992ec,Kuznetsov:1997az} fails for energy
above the sphaleron due to the problems one encounters as the energy
approaches the height of the barrier in systems with many degrees of
freedom.  These problems were studied in detail in the context of a
quantum mechanical model in Ref.~\cite{Bezrukov:2003yf}, where a
regularization technique was suggested to overcome them.

The basic idea in the proposal of Refs.~%
\cite{Rubakov:1992fb,Tinyakov:1992fn,Rubakov:1992ec,Kuznetsov:1997az}
is that, instead of a process with exclusive, two-particle initial
state, one considers a topology changing process with inclusive
initial state characterized by definite energy $E$ and incoming
particle number $N$.  The transition probability $\sigma(E,N)$ can
then be used to provide a bound on the exclusive two-particle
cross-section, while the two-particle transition exponent is obtained
in the limit $\alpha_WN\to0$.

\begin{figure}
  \begin{center}%
\includegraphics{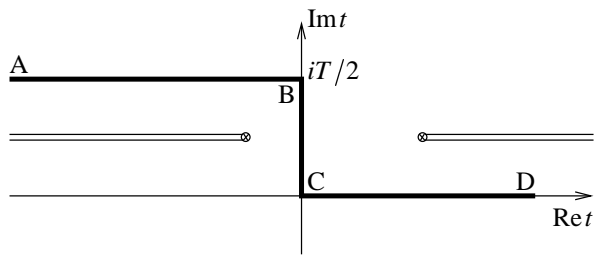}
  \end{center}
  \caption{The contour in complex time plane used in the formulation
    of the boundary value problem \eqref{final_BVP}.  Crossed circles
    represent singularities of the field.  If the field is spherically
    symmetric in space, the singularities closest to imaginary axis
    occur at $r=0$, for other $r$ the singularities generally move to
    larger $|\Real t|$.}
  \label{fig:time_contour}
\end{figure}

The inclusive probability of tunneling from a state with fixed energy
and number of particles is
\begin{equation}\label{sigmaEN}
  \sigma(E,N) =
    \sum_{i,f} |\bra{f} \hat{S} \hat{P}_E \hat{P}_N \ket{i}|^2 \;,
\end{equation}
where $\hat{S}$ is the $S$-matrix, $\hat{P}_{E,N}$ are projectors onto
subspaces of fixed energy $E$ and fixed number of particles $N$, and
the states $\ket{i}$ and $\ket{f}$ are perturbative excitations about
topologically distinct vacua.  This matrix element can be written in
double path integral representation.  For large $N=\tilde{N}/\alpha_W$
and $E=\tilde{E}/\alpha_W$ the path integral can be calculated in the
semiclassical approximation, and this leads to the problem of solving
the equations of motion of the system on a special contour in complex
time plane, which detours around singularities, as shown in
Fig.~\ref{fig:time_contour}.  The presence of branch cut singularities
can be inferred from the following argument.  One notices that, for
energy below the sphaleron energy, if one continues the solution along
a line parallel to the real axis, be this via a forward integration of
the equation of motion from the AB part of the contour or a backward
integration from the CD part of the contour, the field must fall back
to the original topological sector.  On the other hand, by
construction, on the AB and CD parts of the contour the solution must
be in different topological sectors.  Thus the solution must also be
in different topological sectors on the AB part of the contour and on
the negative real axis and, likewise, on the positive real axis and
the continuation of the AB segment to positive time.  This may happen
only if two branch cut singularities exist on the two sides of the BC
part of the contour (see Fig.~\ref{fig:time_contour}).

Eventually the semiclassical approximation produces the following
result ($\varphi$ here stands for all physical fields in the model),
\begin{gather}\label{sigmaFexp}
  \sigma(E,N) \sim
  \exp \left\{ - \frac{4\pi}{\alpha_W} F(\tilde{E},\tilde{N}) \right\} \\
  \frac{4\pi}{\alpha_W} F(\tilde{E},\tilde{N}) =
  2\Imag S_{ABCD}(\varphi) - N\theta - ET -\Real\cB_i
  \;.\nonumber
\end{gather}
Here $S_{ABCD}(\varphi)$ is the action along the time contour, the
parameters $T$ and $\theta$ are Legendre conjugate to $E$ and $N$; the
parameter $T$ is the same as in Fig.~\ref{fig:time_contour}; we will
have to say more about $\theta$ later on.  In what follows we will
usually drop the tilde over the rescaled energy and incoming particle
number, and the overall $1/\alpha_W$ factor, restoring it only in the
final results.  The boundary term
\[
  \cB_i=\frac{1}{2}\int\! d\bk(\fk \fmk\e^{-2i\omega_\bk(T_i-iT/2)}
        -g_{\bk}^* g_{-\bk}^*\e^{2i\omega_\bk(T_i-iT/2)})
\]
is written using frequency components $\fk$ and $g\bkind$ of the field
on the part A of the contour:
\begin{align}
  \varphi({\bm x},t)\big|_{t\to-\infty+iT/2} = \\
  \int \frac{d\bk}{ (2\pi)^{3/2}{\sqrt{2\omega_{\bk}}} }
   \Big( & \fk \e^{-i\omega_\bk(t-iT/2)+i{\bm{kx}}} \nonumber\\
         & + g_{\bk}^* \e^{i\omega_\bk(t-iT/2)-i{\bm{kx}}}  \Big)
  \;. \nonumber
\end{align}
The field $\varphi$ satisfies the field equation
\begin{subequations}\label{final_BVP}
\begin{equation}
  \label{BVP_eq}
  \vpar{S}{\varphi} = 0
\end{equation}
At initial time the frequency components of the solution should
satisfy the following equation (``$\theta$ boundary condition'')
\begin{equation}\label{BVP_bc2}
  \fk = \e^{-\theta} g_{\bk} \;.
\end{equation}
For $\theta$ different from zero this equation implies that the field
must be continued to complex values.  For a complex field, like $\Phi$
in~\eqref{SU2action}, its real and imaginary parts must be continued to
complex values separately.

On the final part of the contour (CD), the field must satisfy the reality
condition
\begin{equation}\label{BVP_real}
  \Imag \dot \varphi(\bm{x},T_f\to\infty) \to 0 \;,\quad
  \Imag \varphi(\bm{x},T_f\to\infty) \to 0
\end{equation}
\end{subequations}
(for complex fields, such as $\Phi$ in \eqref{SU2action}, this means
that both $(\Phi+\Phi^\dagger)/2$ and
$(\Phi-\Phi^\dagger)/2i$ must be real).

Equations
\eqref{BVP_eq}--\eqref{BVP_real} specify the boundary value problem
corresponding to the induced topological transition.

The equations obtained by variation over the auxiliary parameters $T$ and
$\theta$ are
\begin{align}
  \label{BVP_E}
  E &= \int d\bk\, \omega\bkind \fk\gsk \\
  \label{BVP_N}
  N &= \int d\bk\, \fk\gsk \;.
\end{align}
These equations indirectly fix values of $T$ and $\theta$ for given
energy and number of particles.  Alternatively, one can fix $T$ and
$\theta$, solve the boundary value problem \eqref{final_BVP} and
obtain the corresponding values of $E$ and $N$ using \eqref{BVP_E} and
\eqref{BVP_N}.  This is especially convenient in numerical
calculations.

The interpretation of the solutions to the boundary value problem
(\ref{final_BVP}) is as follows.  On the part CD of the contour, the
saddle-point field is real asymptotically; it describes the evolution
of the system after tunneling.  On the contrary, it follows from
boundary conditions (\ref{BVP_bc2}) that in the initial asymptotic
region the saddle-point field is complex whenever $\theta\neq 0$.
Thus, the initial state which maximizes the probability
(\ref{sigmaFexp}) is not described by a real classical field, i.e.\
this stage of the evolution is essentially quantum even at $N\sim
1/\alpha_W$.

There is a subtle point concerning the boundary condition
\eqref{BVP_real}.  It can be satisfied in two different ways.  Either
the solution is exactly real on the whole CD part of the contour and
is close to vacuum, or it has an exponentially decaying imaginary part
and approaches the sphaleron along the complexified unstable
direction.  This subtlety is important for the analysis at high
energies (section~\ref{sec:reg}), $E\gtrsim E_\mathrm{sph}$.

\begin{figure}
  \begin{center}
    \includegraphics{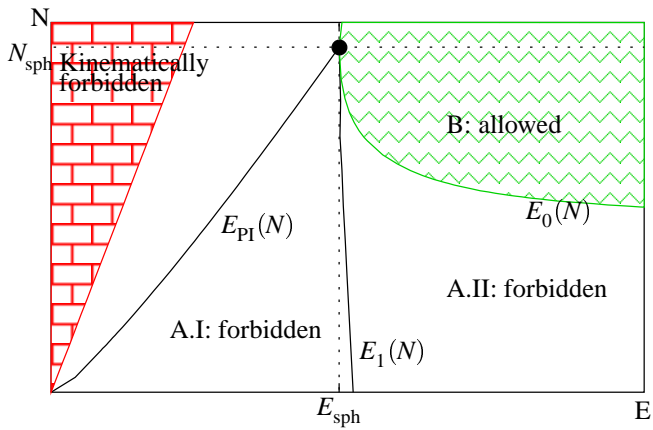}
    \caption{Regions in the $E$--$N$ plane.}
    \label{fig:regions}
  \end{center}
\end{figure}

The solutions to the boundary value problem can be found numerically
for different values of $E$ and $N$.
In this paper we study solutions that have spherical symmetry in
space.  One expects that these are most important for large enough
$N$; perturbative calculations about the instanton suggest that
spatial spherical symmetry is relevant at relatively low energies and
all $N$.  We do not have a convincing argument in favor of spherical
symmetry for few particle collisions at very high energies; in any
case, our results as they stand, are valid for s-wave scattering.

Our numerical analysis shows that
the $E-N$ plane is divided into
several different regions (see Fig.~\ref{fig:regions}).  Values of
$E<N\cdot\min(M_W,M_H)$ are trivially excluded by kinematics.  For
relatively low energies (region A) the transitions between the
topologically distinct vacua can occur only via tunneling.  At the
sphaleron energy $E_\mathrm{sph}$ the situation changes.  A slight
excitation of the sphaleron along the unstable direction gives origin
to a solution of the classical equations of motion which evolves
towards different topological sectors at large negative and positive
times.  Since the sphaleron has exactly one negative mode, there is
only one infinitesimal deformation of this type, and thus the
corresponding solution has definite number of particles
$N_{\mathrm{sph}}$ in the initial state.  At higher energies one may
add excitations of the positive modes above the sphaleron to obtain
over-barrier solutions with different, and, in particular, smaller
initial particle number.  These solutions belong to the domain of
classically allowed transitions (region B in Fig.~\ref{fig:regions}),
where the topology changing processes are unsuppressed.  The boundary
between region A and region B corresponds to configurations staying
for an infinite time close to the sphaleron (since there are no bound
states in the sphaleron background~\cite{Akiba:1989xu}, all excitations
about the sphaleron fly away at finite time, so the field relaxes to
the sphaleron solution).

In the classically forbidden region A there is a special family of
solutions, corresponding to $\theta=0$ in the boundary value problem.
These are represented by the line $E_\mathrm{PI}(N)$ in
Fig.~\ref{fig:regions}.  In this case, the boundary condition
(\ref{BVP_bc2}) reduces to the reality condition imposed at $\Imag t
=T/2$.  The solution to the resulting boundary value problem is the
periodic instanton of Ref.~\cite{Khlebnikov:1991th}.  The periodic
instanton is a real periodic solution to the Euclidean field equations
with period $T$ and turning points at $t=0$ and $t=iT/2$
$(\mathrm{mod }T)$.  When analytically continued in the Minkowskian
direction through the turning points, the periodic instanton is real
on the lines $\Imag t =0$ and $\Imag t = T/2$ and therefore satisfies
the boundary value problem (\ref{final_BVP}) with $\theta=0$.  Like
any other solution linearizing at large negative times at part A of
the contour of Fig.~\ref{fig:time_contour}, the periodic instanton has
a certain number of incoming particles, Eq.~\eq{BVP_N}.  For given
energy $E$ below the sphaleron, this number is such that the
suppression exponent $F(E,N)$ has a minimum, i.e., the transition
occurs at maximum rate.

The classically forbidden region A is further subdivided into two
regions.  For low energies (region A.I) the system is close to the
vacuum on the final part of the evolution, so the boundary condition
\eqref{BVP_real} leads to the exact reality of the fields on the part
CD of the time contour.  At energies higher than the sphaleron energy
(precisely, on the right of the line $E_1(N)$) the system ends up
close to the sphaleron (with extra outgoing waves in the sphaleron
background).  In this case Eq.~\eqref{BVP_real} is truly asymptotic.
So, the system tunnels ``on top'' of the barrier, creating an unstable
sphaleron configuration, which then decays with probability of order 1
to any of the two neighboring vacua.  This situation is realized in
the region A.II.  This new qualitative feature of the tunneling at
high energies emerges from the existence of the bifurcation of the
solutions and is not seen in any order of perturbative expansion
around the instanton.  Non-perturbative approaches, however, capture
this feature (see also Ref.~\cite{Voloshin:1994dk} for similar results
in the context of false vacuum decay).

\begin{figure}
  \centerline{\includegraphics[width=\columnwidth]{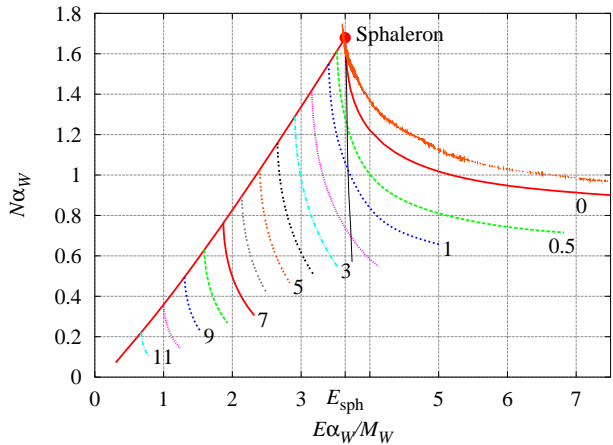}}
  \caption{Lines of $F(E,N)=\const$.  Lines are labeled by the values
    of the suppression exponent $-\alpha_W\log\sigma=4\pi F$.
    Diagonal line directed from the sphaleron towards the origin is
    the line of periodic instantons.  Energy $E$ is in units of
    $M_W/\alpha_W$, number of particles $N$ is in units of
    $1/\alpha_W$.  The line labeled by $0$ ($F=0$) is the boundary of
    the classically allowed region $E=E_0(N)$. The ``fuzzy'' line
    represents the approximate boundary of the classically allowed
    region found in over-barrier calculations of
    Ref.~\cite{Rebbi:1996zx}.}
  \label{fig:ne_f}
\end{figure}

\begin{figure}
  \centerline{\includegraphics[width=\columnwidth]{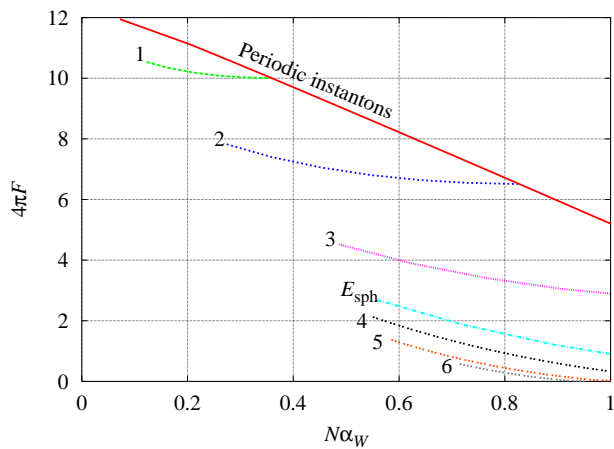}}
  \caption{Dependence of the suppression exponent on the number of
  particles $N$ for different energies.  Numbers near the curves are
  energies in units of $M_W/\alpha_W$.}
  \label{fig:fn_e}
\end{figure}

Our numerical results for the suppression exponent in the whole
classically forbidden region are presented in Figs.~\ref{fig:ne_f},
\ref{fig:fn_e}.  The almost vertical line in Fig.~\ref{fig:ne_f}
separates the two regions (denoted by A.I and A.II in our earlier
discussion) where the tunneling process assumes characteristically
distinct features.  It also represents the frontier beyond which
numerical calculations based on a straightforward implementation of
the method of Refs.~%
\cite{Rubakov:1992fb,Tinyakov:1992fn,Rubakov:1992ec,Kuznetsov:1997az}
appear to fail.  It is clear from Fig.~\ref{fig:ne_f} that our
improved numerical technique can go well beyond that frontier.
Reference~\cite{Bezrukov:2003zn} presents a comparison between our
results and the analytic predictions for the suppression exponent
$F(E,N)$ in the limit of small energy.  The two are in remarkable
agreement which provides a gratifying check of the numerical
calculations.

Another interesting comparison can be made with the results of
Ref.~\cite{Rebbi:1996zx}, where the real-time overbarrier solutions
close to the boundary of the classically allowed region were searched
via Monte-Carlo techniques\footnote{In Ref.~\cite{Rebbi:1996zx} the
coupling constant $\lambda$ was chosen $0.1$, while we use
$\lambda=0.125$.  We performed a set of calculations for
$\lambda=0.1$.  The dependence on $\lambda$ is so weak, that the
difference for the results would be invisible in the graph.  Much
larger discrepancies appear because of the different lattice parameters used in
the two calculations. (In Ref.~\cite{Rebbi:1996zx}, having only to solve 
for the real time evolution of the fields, it was possible to use a larger 
lattice and a finer lattice spacing than in the present calculations.)}.
In this way, an approximation (which, at the same
time, is an upper bound) for the boundary of the classically allowed
region was obtained.  It is seen that the results of
Ref.~\cite{Rebbi:1996zx} are reasonably close to the boundary
$E_0(N)$, found in our calculations.

\begin{figure}
  \centerline{\includegraphics[width=\columnwidth]{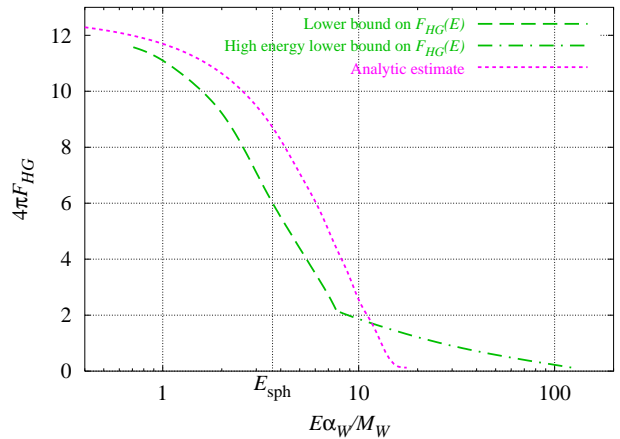}}
  \caption{Lower bound on the suppression exponent for two-particle
    collisions, dashed and dashed-dotted lines.  Dotted line is the
    estimate of Refs.~\cite{Ringwald:2002sw,Ringwald:2003px}.}
  \label{fig:hg}
\end{figure}

\begin{figure}
  \centerline{\includegraphics[width=\columnwidth]{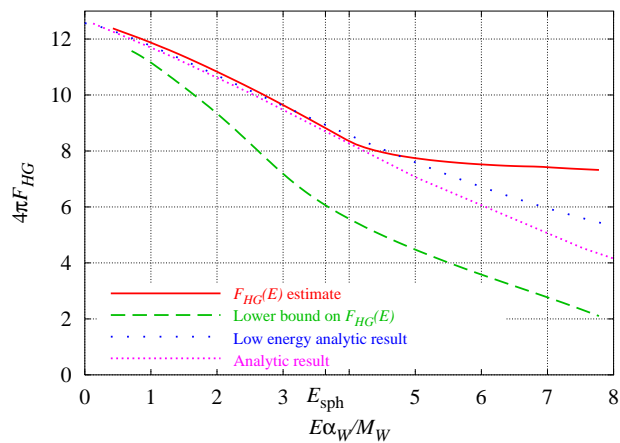}}
  \caption{Estimate of the suppression exponent for two-particle
    collisions $F_{HG}(E)$ (solid line), lower bound on $F_{HG}(E)$
    (dashed line), low energy analytic prediction~\eqref{HGanalytic}
    (rare dotted line) and analytic estimate of
    Refs.~\cite{Ringwald:2002sw,Ringwald:2003px} (dotted line).}
  \label{fig:hg_low}
\end{figure}

Our results by themselves do not reach the physically interesting
$N=0$ limit (corresponding to particle collisions).  Studying lower
$N$ in numerical calculations would need lattices with larger number
of lattice points and would require quite substantial amounts of time
even on powerful present day supercomputers.  Therefore, some
extrapolation must still be used to get insight on the suppression
factor for actual particle collisions.  As we seek such
extrapolations, we notice first that it is quite straightforward to
obtain a lower bound on the suppression exponent $F$.  Insofar as
$\theta$ increases as $N\to0$, and $(4\pi)\partial F/\partial
N=-\theta$, by simply continuing $F$ with a linear function of $N$ for
each energy one obtains a lower bound on $F$.  This bound is shown in
Figs.~\ref{fig:hg}, \ref{fig:hg_low}, dashed line.  It indicates that
up to the energy $8M_W/\alpha_W\simeq 20\TeV$ the suppression is still
high: the suppression factor is smaller than $\e^{-60}\sim10^{-26}$
for $\alpha_W\sim1/30$.

For very high energies a bound may be constructed by exploiting the
observation that the lines of constant $F$ in $E-N$ plane have
positive curvature (see Fig.~\ref{fig:ne_f}).  So, by extrapolating
these lines linearly to $N=0$ one obtains another lower bound on the
suppression exponent $F(E,N=0)$.  This bound is displayed in
Fig.~\ref{fig:hg}, dashed-dotted line.  One can see that exponential
suppression continues up to an energy of at least $250\TeV$.

One may also attempt to estimate the function $F(E)$ itself.  As we
discuss in Section~\ref{sec:num_results}, a good estimate is obtained
by extrapolating, instead of $F(E,N)$, the function $T(N)$ at fixed
energy, as $T(N)$ is approximately linear in $N$.  Up to the sphaleron
energy, the estimate obtained in this way is close to the one loop
analytic result~%
\cite{Khoze:1991bm,Arnold:1991cx,Diakonov:LINPSchool1991,Mueller:1991fa},
which gives three terms in the low-energy expansion,
\begin{equation}\label{HGanalytic}
  \frac{4\pi}{\alpha_W}F(E) = \frac{4\pi}{\alpha_W}\left[
    1-\frac{9}{8}\left(\frac{E}{E_0}\right)^{4/3}
    +\frac{9}{16}\left(\frac{E}{E_0}\right)^2
  \right],
\end{equation}
where $E_0=\sqrt{6}\pi M_W/\alpha_W$.  Below the sphaleron, our
estimate is also consistent with the analytic estimate of
Refs.~\cite{Ringwald:2002sw,Ringwald:2003px}.  On the other hand, the
behavior of $F_{HG}(E)$ changes dramatically at $E\gtrsim
E_\mathrm{sph}$.  We attribute this to the change in the tunneling
behavior---at $E\gtrsim E_\mathrm{sph}$ the system tunnels ``on top of
the barrier''.  Our numerical data show that the suppression exponent
$F_{HG}(E)$ flattens out, and topology changing processes are in fact
much heavier suppressed at $E\gtrsim E_\mathrm{sph}$ as compared to
the estimate~\eqref{HGanalytic} and the estimate of
Refs.~\cite{Ringwald:2002sw,Ringwald:2003px}.  We show our estimate,
together with analytical estimates and our lower bound, in
Fig.~\ref{fig:hg_low}.

It is worth noting that similar effects of dramatic change of the
behavior of the system at high energies were observed in lattice
calculations of instanton distribution in QCD
in~\cite{Ringwald:1999ze,Schrempp:2002kd}.

Thus, our numerical results, albeit covering a limited range of
energies and initial particle numbers, enable us to obtain both lower
bound for and actual estimate of the suppression exponent for the
topology changing two-particle cross-section in the electroweak theory
well above the sphaleron energy.  This cross section remains
exponentially suppressed up to very high energies of at least
$250\TeV$.  In fact, the energy, if any, at which the exponential
suppression disappears, is most likely much higher, as suggested by
comparison of our lower bound and actual estimate at energies
exceeding significantly $E_\mathrm{sph}$, see Fig.~\ref{fig:hg_low}.

\section{The method of RST}
\label{sec:rst}

\subsection{General formulation}
\label{sec:rst_general}

The quantity we wish to calculate is $\sigma(E,N)$, the probability
of transition from a state with fixed energy $E$ and number of
particles $N$ about one vacuum to \emph{any} state about another
vacuum.  The method of semiclassical calculation of this inclusive
multiparticle probability was formulated in Refs.~%
\cite{Rubakov:1992fb,Tinyakov:1992fn,Rubakov:1992ec,Kuznetsov:1997az}.
We call it the RST method for brevity, and here we review this
prescription in brief.

\begin{figure}
  \begin{center}
\includegraphics{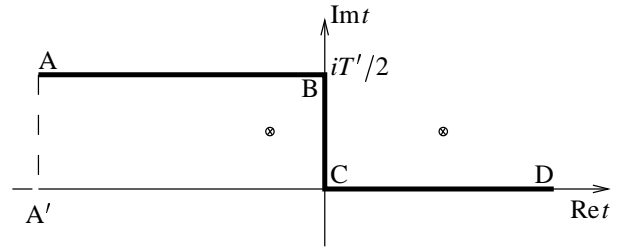}
  \end{center}
  \caption{The contour used to derive the boundary value problem.}
  \label{fig:time_contour2}
\end{figure}

The inclusive multiparticle probability \eqref{sigmaEN} can be written
in functional integral form, where the semiclassical
approximation is equivalent to the saddle-point integration.  The
double path integral representation for $\sigma(E,N)$
reads~\cite{Rubakov:1992fb}
\begin{widetext}
\begin{multline}\label{PIsigmaE,N}
  \sigma(E,N) = \int\! d\vartheta d\mathcal{T} da\bkind da^*_\bk
    db\bkind db^*_{\bk} d\varphi(x) d\varphi'(x)\,
    \exp \Bigl\{ -iN\vartheta
  -iE\mathcal{T}
    -\int d\bm{k} a\bkind a^*_{\bk} \e^{-i\vartheta - i\omega_\bk\mathcal{T}}
    -\int d\bm{k} b\bkind b^*_{\bk}
\\
  + B_i(a_{\bk},\varphi_i) + B_f(b^*_{\bk},\varphi_f)
    + B^*_i(a^*_{-\bk},\varphi_i') 
  + B^*_f(b_{-\bk},\varphi_f')
    + i S(\varphi) -i S(\varphi') \Bigr\} \;.
\end{multline}
Here $\varphi$ stands for all physical fields of the theory.
The boundary terms $B_i$ and $B_f$ are
\begin{eqnarray}
  B_i(a_{\bk},\varphi_i) &=&\half  \int\! d\bm{k}\, \Big[
    - \omega_{\bk} \varphi_i(\bm{k})\varphi_i(-\bm{k})
    -  a_{\bk} a_{-\bk} e^{-2i\omega_\bk T_i}
    + 2\sqrt{2\omega_{\bk}}\, e^{-i\omega_\bk T_i}
    a_{\bk} \varphi_i(\bm{k}) \Big] \; ,
\label{Bi,f}\\
  B_f(b^*_{\bk},\varphi_f) &=&  \half \int\! d\bm{k}\, \Big[
    -  \omega_{\bk} \varphi_f(\bm{k}) \varphi_f(-\bm{k})
    -  b^*_{\bk} b^*_{-\bk}e^{2i\omega_\bk T_f}
    + 2\sqrt{2\omega_{\bk}}e^{i\omega_\bk T_f}
    b^*_{\bk} \varphi_f(-\bm{k}) \Big] \; ,
  \nonumber
\end{eqnarray}
\end{widetext}
where $\varphi_{i,f}(\bm{k})$ are the spatial Fourier transforms of the
field at initial and final times $T_i$ and $T_f$, respectively.  The
limit $T_{i,f}\to\mp\infty$ is assumed at the end of the calculation.
The complex integration variables $a_{\bk}$ and $b^*_{\bk}$ come from
the coherent state representation of initial and final states; they are
the classical counterparts of annihilation and creation operators.  The
integration over these variables implements the summation over
initial and final states in \eq{sigmaEN}.  The functional integrals
over $\varphi(x)$ and $\varphi'(x)$ come from the amplitude and
complex conjugate amplitude, respectively.  The integrations include
the boundary values $\varphi_{i,f}$ and $\varphi'_{i,f}$.  Integration
over $\mathcal{T}$ and $\vartheta$ serve to project onto the
subspaces of fixed $E$ and $N$, respectively.

The integral~\eqref{PIsigmaE,N} can be evaluated in the saddle point
approximation, as long as the exponent is proportional to
$1/\alpha_W$, implicitly present in the expression, and
$N,E\sim1/\alpha_W$.

Let us now discuss the saddle-point equations for the integral
(\ref{PIsigmaE,N}).  We will see that these equations reduce to a
certain boundary value problem for the fields $\varphi$ and
$\varphi'$.  The variables $a\bkind$, $a^*_{\bk}$, $b\bkind$ and
$b^*_{\bk}$ enter the exponent quadratically and can be integrated
out, yielding
\begin{widetext}
\begin{multline}\label{smallPIsigmaEN}
  \sigma(E,N) = \int d\vartheta d\mathcal{T} d\varphi(x) d\varphi'(x)
    \prod\limits_{\bk} \delta (\varphi_f(\bm{k}) - \varphi'_f(\bm{k}))
   \times \exp \biggl\{
    -iN\vartheta -iE\mathcal{T} + i S(\varphi) -i S(\varphi') \\
   - \half \int d\bm{k} \frac{ \omega_{\bk}}{1 - \gamma_{\bk}^2 }
    \Bigl( (1+\gamma_{\bk}^2)
    [\varphi_i(\bm{k}) \varphi_i(-\bm{k}) 
   + \varphi'_i(\bm{k}) \varphi'_i(-\bm{k})]
    - 4 \gamma_{\bk} \varphi_i(\bm{k}) \varphi'_i(-\bm{k})
    \Bigr)
    \biggl\} \;,
\end{multline}
\end{widetext}
where
\[
  \gamma_{\bk} = \e^{i\vartheta+i\omega_\bk \mathcal{T}}\;.
\]
An important feature of the representation (\ref{smallPIsigmaEN}) is
that the exponent in the r.h.s.\ contains only the action and the
boundary terms.  Thus, the discretization of this exponent is
relatively straightforward.

Let us turn to the saddle point equations.  Varying the exponent with
respect to the fields $\varphi(x)$ and $\varphi'(x)$ we find
\begin{equation}\label{field-eqs}
  \vpar{S}{\varphi} = \vpar{S}{\varphi'} =0 \;,
\end{equation}
i.e.\ the usual field equations.  The boundary conditions for these
equations come from the variation with respect to the boundary values
of the fields.  At $t=T_f$, because of the $\delta$--function, the
variations are subject to the constraint $\delta\varphi_f(\bm{x}) =
\delta\varphi'_f(\bm{x})$ (at $T_f\to\infty$).  Since $\delta S
/\delta \varphi(T_f,\bm{x}) = \dot\varphi(T_f,\bm{x})$ we obtain
\begin{eqnarray}  \label{bc-1}
  \dot\varphi(T_f,\bm{x}) & = & \dot\varphi'(T_f,\bm{x}) \;,
  \\
  \varphi(T_f,\bm{x}) & = & \varphi'(T_f,\bm{x})\;.
  \nonumber
\end{eqnarray}
Thus, in the final asymptotic region the saddle-point fields $\varphi$
and $\varphi'$ coincide.

The variation with respect to $\varphi_i$ and $\varphi'_i$ leads to
two equations which can be written in the following form,
\begin{eqnarray*}
  i\dot\varphi_i(\bm{k}) + \omega_{\bk} \varphi_i(\bm{k}) & = &
    \gamma_{\bk} \left( i\dot\varphi'_i(\bm{k})
    + \omega_{\bk} \varphi'_i(\bm{k}) \right) \;,
  \\
  -i\dot\varphi_i(\bm{k}) + \omega_{\bk} \varphi_i(\bm{k}) & = &  
    \frac{1}{\gamma_{\bk}} \left( -i\dot\varphi'_i(\bm{k}) + \omega_{\bk}
    \varphi'_i(\bm{k}) \right) \;.
  \nonumber
\end{eqnarray*}
These initial boundary conditions simplify when written in
terms of frequency components.  In the initial asymptotic region
($t\to-\infty$), where $\varphi$ and $\varphi'$ are free fields, we
can write
\begin{align}
  \varphi(x) &=\!
    \int\!\!{\textstyle\frac{d\bk}{(2\pi)^{\frac{3}{2}} \sqrt{2\omega_k\!}}}\!
    \left\{ f\bkind \e^{-i\omega_\bk t+i{\bm{kx}}}
      + g_{\bk}^* \e^{i\omega_\bk t-i{\bm{kx}}}
    \right\}\! ,
  \label{fkdef}  \\
  \varphi'(x) &=\!
    \int\!\!{\textstyle\frac{d\bk}{(2\pi)^{\frac{3}{2}} \sqrt{2\omega_k\!}}}\!
    \left\{ f'_{\bk} \e^{-i\omega_\bk t+i{\bm{kx}}}
      + {g'_\bk}^* \e^{i\omega_\bk t-i{\bm{kx}}}
    \right\}\! .
  \nonumber
\end{align}
Then the initial boundary conditions become
\begin{eqnarray}
  f\bkind &=& \gamma_{\bk} f'_{\bk} \;,
  \nonumber\\
  g^*_{\bk} &=& {1\over \gamma_{\bk}} {g'}^*_{\bk} \;,
  \label{simplebc-2}
\end{eqnarray}
Finally, there are two saddle-point equations which come from the
variation of the exponent in \eq{smallPIsigmaEN} with respect to
$\vartheta$ and $\mathcal{T}$.  These equations determine the
saddle-point values of $\vartheta$ and $\mathcal{T}$ as functions of
$E$ and $N$.  In terms of frequency components $f_\bk$ and $g_\bk$
they read (after using boundary conditions \eqref{simplebc-2})
\begin{eqnarray}
  E &=& \int d\bm{k} \omega\bkind f\bkind g^*_{\bk},
  \label{Espec2} \\
  N &=& \int d\bm{k} f\bkind g^*_{\bk}.
  \label{Nspec2}
\end{eqnarray}
One may recognize the usual expressions for the energy and the number
of particles contained in the free classical field, $n\bkind=f\bkind
g^*_{\bk}$ being the occupation number in the mode with spatial
momentum $\bm{k}$.

The field $\varphi'(x)$ originates from the complex conjugate
amplitude.  This suggests that its saddle point value is complex
conjugate to that of $\varphi(x)$.  Indeed, the Ansatz
\begin{equation*}
  [\varphi(t,\bm{x})]^* =\varphi'(t,\bm{x})
\end{equation*}
is compatible with the boundary value
problem~\eqref{field-eqs}--\eqref{Nspec2}.  Then the saddle point
values of $\mathcal{T}$ and $\vartheta$ are pure imaginary
\[ 
  \mathcal{T}=iT\;,\qquad \vartheta=i\theta \;,
\]
provided the initial energy~\eqref{Espec2} and particle
number~\eqref{Nspec2} are real.  The boundary conditions~\eqref{bc-1}
imply then that the field $\varphi$ is real asymptotically at final
time
\[
  \Imag \dot \varphi(T_f,\bm{x}) \to0
  \;,\quad
  \Imag \varphi(T_f,\bm{x}) \to 0 
     \quad \text{for }T_f\to+\infty\;,
\]
while Eq.~\eqref{simplebc-2} relates the positive and negative
frequency components of the field $\varphi$ in the initial asymptotic
region
\[
  f_\bk = \gamma_\bk g_\bk\;,
\]
where
\begin{equation}\label{gamma}
  \gamma_{\bk} = \e^{-\theta - \omega_kT} \;,
\end{equation}
Until now, the initial time $T_i$ was real.  However, it is convenient
to reformulate the boundary value problem directly in terms of the
fields on the contour ABCD, at which the initial time has imaginary
part $\Imag T_i=T'/2$ (see Fig.~\ref{fig:time_contour2}).  The
analytical continuation in the initial asymptotic region can be done
explicitly by means of Eqs.~(\ref{fkdef}).  In
Eqs.~(\ref{simplebc-2})--(\ref{Nspec2}) this continuation results in
the substitution of $\gamma_{\bk}$ by
\[
  \gamma_\bk = \e^{-\theta-\omega_\bk(T-T')}.
\]
The simplest boundary conditions are obtained in the case
when the contour height in imaginary time $T'$ is equal to the
parameter $T$, leading to $\bk$-independent $\gamma$
\[
  \gamma=\e^{-\theta} \;.
\]
In this case one arrives at the boundary condition~\eqref{BVP_bc2} and
the contour ABCD with height $T/2$ shown in
Fig.~\ref{fig:time_contour}.  This formulation will be used in most
cases.  Then the boundary value
problem~\eqref{field-eqs}--\eqref{Nspec2} is equivalent\footnote{The
  boundary term in~\eqref{sigmaFexp} is obtained from the boundary terms
  in~\eqref{smallPIsigmaEN} by making use of the $\theta$ boundary
  conditions~\eqref{BVP_bc2}.}  to~\eqref{final_BVP}--\eqref{BVP_N}.
This is the boundary value problem we solve numerically in the present
paper.

Let us discuss some subtle points of this boundary value problem.
First, one notices that the condition of asymptotic reality
\eqref{BVP_real} does not always coincide with the condition of
reality at finite time.  Of course, if the solution approaches the
vacuum on the part CD of the contour, the asymptotic reality condition
\eqref{BVP_real} implies that the solution is real at any
\emph{finite} positive $t$.  Indeed, at large enough time the system
evolves in that case in the linear regime, so the condition
\eqref{BVP_real} means that all physical modes should be real.  Due to
the equations of motion the fields are then real on the entire CD--part of
the contour.  This situation corresponds to the transition directly to
the neighboring vacuum.  However, the situation can be drastically
different if the solution on the final part of the time contour
remains in the interaction region, i.e.\ close to the sphaleron.
Since one of the excitations about the sphaleron is unstable, there
may exist solutions which approach the sphaleron \emph{exponentially}
along the complexified unstable direction.  In that case the solution
may be complex at any finite time, and become real only
asymptotically, as $t\to+\infty$.  Such solution corresponds to
tunneling to the sphaleron; afterwards the system rolls down
classically to the correct vacuum (with probability of order $1$,
inessential for the tunneling exponent $F$).  We will see in
section~\ref{sec:reg} that the situation of this sort indeed takes
place at high energies $E\gtrsim E_\mathrm{sph}$.

Second, the initial boundary conditions \eqref{simplebc-2} (imposed on
the real time axis) mean, that $\varphi$ and $\varphi'=\varphi^*$ are
different at large negative time, while at large positive time they
coincide because of the condition~\eqref{bc-1}.  For solutions ending
in the vacuum at positive time (so that the fields are exactly real at
finite $t>0$), this means that there should exist a branch point in
the complex time plane: the contour in Fig.~\ref{fig:time_contour}
winds around this point and cannot be deformed to the real time axis.
This argument \emph{does not} work for solutions ending on the
sphaleron at $t\to+\infty$, so branch points between the AB--part of
the contour and the real time axis may be absent.  We have found that
this is indeed the case at high energies
(cf.\ Ref.~\cite{Bezrukov:2003yf}).

\subsection{Reduction to spherically symmetric configurations}
\label{sec:spher}

Here we consider spherically symmetric
configurations~\cite{Ratra:1988dp} of the $SU(2)$--Higgs theory.  The
reason is that one can entertain the expectation that the most
important tunneling configurations possess maximum spatial symmetry.
On the other hand, without the simplification provided by
spherical symmetry the computational cost of the numerical analysis 
would be prohibitive.

In the spherically symmetric \emph{Ansatz} the original fields are
expressed in terms of six real two-dimensional fields $a_0$, $a_1$,
$\alpha$, $\beta$, $\mu$ and $\nu$ as follows
\begin{align}
  A_0(\bm{x},t) &= \half a_0(r,t) \bm{\sigma\cdot n}
  \nonumber\\
  A_i(\bm{x},t) &= \half\bigg[a_1(r,t)\bm{\sigma\cdot n}n_i
                   +\frac{\alpha(r,t)}{r}(\sigma_i
                                          -\bm{\sigma\cdot n}n_i)
  \nonumber\\
                &\hphantom{=\half\bigg[}
                   +\frac{1+\beta(r,t)}{r}\epsilon_{ijk}n_j\sigma_k
                   \bigg]
  \label{ansatz}\\
  \Phi(\bm{x},t) &= [\mu(r,t)+i\nu(r,t)\bm{\sigma\cdot n}]\xi \;,
  \nonumber
\end{align}
where $\bm{n}$ is the unit three-vector in the radial direction and
$\xi$ is an arbitrary constant two-component complex unit column.
This Ansatz is symmetric under spatial rotations complemented by
appropriate rotations in the gauge group and custodial global symmetry
transformations.  The action \eqref{SU2action} expressed in terms of
the new fields becomes
\begin{widetext}
\begin{multline}\label{2dim_action}
  S = \int\! dt \int_0^\infty\! dr
      \left[\vphantom{\half}\right. 
        \frac{1}{4}r^2f_{\mu\nu}f_{\mu\nu}
        + (\bar D_\mu \bar \chi)D_\mu \chi
        + r^2 (\bar D_\mu \bar\phi)D_\mu\phi \\
      -\frac{1}{2 r^2}\left( ~\bar\chi\chi-1\right)^2
        -\frac{1}{2}(\bar\chi\chi + 1)\bar\phi\phi 
      \left.\vphantom{\half}
        -\frac{i}{2} \bar\chi \phi^2+\frac{i}{2} \chi \bar\phi^2
        -{\lambda}  r^2 (\bar\phi\phi- 1)^2 \right]
\end{multline}
\end{widetext}
where the indices $\mu$, $\nu$ run from 0 to 1 and
\begin{subequations}\label{defns}
\begin{align}
  f_{\mu\nu}&=\partial_\mu a_\nu - \partial_\nu a_\mu \\
       \chi &=\alpha+i\beta&
   \bar\chi &= \alpha - i\beta\\
       \phi &= \mu + i \nu&
   \bar\phi &= \mu - i \nu\\
  D_\mu\chi &= (\partial_\mu-i a_\mu)\chi&
  \bar D_\mu\bar\chi &= (\partial_\mu + i a_\mu)\bar\chi\\
  D_\mu \phi &= (\partial_\mu- \frac{i}{2} a_\mu)\phi&
  \bar D_\mu \bar\phi &= (\partial_\mu + \frac{i}{2} a_\mu)\bar\phi
  \;.
\end{align}
\end{subequations}
Note that the overbar on $\phi$, $\chi$ and $D_{\mu}$ denotes changing
$i \to -i$ in the definitions (\ref{defns}) above, which is the same
as complex conjugation \emph{only} if the six fields $a_{\mu}$,
$\alpha$, $\beta$, $\mu$ and $\nu$ are real.  In the boundary value
problem \eqref{final_BVP} these fields become complex and overbar no
longer corresponds to normal complex conjugation.

The equations of motion obtained from \eqref{2dim_action} are
\begin{subequations}\label{2dim_eqs}
\begin{gather}
  \label{gauss_law_eq}
  \partial_1(r^2f_{01}) = i[\chi\bar D_0\bar\chi-\bar\chi D_0\chi]
    +\frac{i}{2}r^2[\phi\bar D_0\bar\phi-\bar\phi D_0\phi]
    \\
  \label{a_eq}
  \partial_0(r^2f_{01}) = i[\chi\bar D_1\bar\chi-\bar\chi D_1\chi]
    +\frac{i}{2}r^2[\phi\bar D_1\bar\phi-\bar\phi D_1\phi]
    \\
  \label{chi_eq}
  \left[\Dm\Dm+\frac{1}{r^2}(\bar\chi\chi-1)
        +\half\bar\phi\phi\right]\chi = -\frac{i}{2}\phi^2 \\
  \label{chibar_eq}
  \left[\bar\Dm\bar\Dm+\frac{1}{r^2}(\bar\chi\chi-1)
        +\half\bar\phi\phi\right]\bar\chi =
     -\frac{i}{2}\bar{\phi}^2 \\
  \label{phi_eq}
  \left[\Dm r^2\Dm+\half(\bar\chi\chi+1)
        +2\lambda r^2(\bar\phi\phi-1)\right]\phi = 
    i\chi\bar\phi \\
  \label{phibar_eq}
  \left[\bar\Dm r^2\bar\Dm+\half(\bar\chi\chi+1)
        +2\lambda r^2(\bar\phi\phi-1)\right]\bar\phi = 
    i\bar\chi\phi \;.
\end{gather}
\end{subequations}
Equation \eqref{gauss_law_eq} is of the first order in time---it is 
Gauss' law.

The spherical Ansatz~(\ref{ansatz}) has a residual $U(1)$ gauge
invariance
\begin{subequations}\label{gaugexform}
\begin{align}
  a_\mu &\to a_\mu + \partial_\mu \Omega \\
  \chi  &\to \e^{i \Omega} \chi \\
  \phi  &\to \e^{i \Omega/2} \phi  \;,
\end{align}
\end{subequations}
with gauge function $\Omega(r,t)$.  The complex ``scalar'' fields
$\chi$ and $\phi$ have $U(1)$ charges $1$ and $1/2$ respectively.
$a_{\mu}$ is the $U(1)$ gauge field, $f_{\mu\nu}$ is the field
strength tensor, and $D_{\mu}$ in \eqref{defns} is the covariant
derivative.  The residual $U(1)$ gauge invariance must be fixed when
solving the equations numerically.  We choose the temporal gauge
$a_0=0$.  In this gauge, if Gauss' law is obeyed at some moment of
time, the other five equations guarantee that it is obeyed at any
time.  This means, in fact, that one of the equations is redundant,
and one of the fields is not physical---it can be expressed in terms
of the other four fields and their derivatives using Gauss' law.
However, numerically it is easier to solve five second order equations
of motion imposing Gauss' law as one of the boundary conditions.
Also, in the $a_0=0$ gauge, there remains a gauge freedom with time
independent gauge function, and this should also be fixed by boundary
conditions.

The trivial space-independent vacuum of the model is
\begin{equation}\label{triv_vac}
  \chi_{\mathrm{vac}}=-i\;,\quad
  \phi_{\mathrm{vac}}=\pm 1\;,\quad
  a_{1\,\mathrm{vac}}=0\;.
\end{equation}
Other vacua are obtained from the trivial one by the gauge
transformations
\begin{subequations}\label{gaugevac}
\begin{align}
  a_{\mu\,\mathrm{vac}} &= \dm\Omega \\
  \chi_\mathrm{vac}     &= -i\e^{i\Omega} \\
  \phi_\mathrm{vac}  &= \pm\e^{i\Omega/2} \;.
\end{align}
\end{subequations}
By regularity, $\Omega$ should be zero at the origin.  Vacua with
different winding numbers correspond to $\Omega\to2n\pi$ as
$r\to\infty$.  For such values of $\Omega$, the fields of the original
four-dimensional model are constant at spatial infinity, and this is
the standard choice.  It allows for a simple description of the
topological properties of vacua: since the sphere $S^2$ at spatial
infinity is mapped to one point in field space, one can compactify the
space to $S^3$ and consider mappings $S^3\to SU(2)$, corresponding to
pure gauge field configurations.

One can also make other choice of gauge transformation function
$\Omega(r)$ at spatial infinity (as long as the fields are pure gauge
and constant in time there).  In our case it is convenient to set
$\Omega\to(2n-1)\pi$ at $r\to\infty$.  This choice, called ``periodic
instanton gauge'' in this paper, in the original 4-dimensional theory
corresponds to mapping of the sphere $S^2$ at spatial infinity onto
the equatorial sphere $S^2$ of the $SU(2)$ gauge group, parameterizing
the pure gauge field configuration.  This behavior of $\Omega$ is
equivalent to the requirement that the fields satisfy the following
boundary conditions at $r=0$ and $r=\infty$,
\begin{align}
  \chi|_{r\to0}                   &\to -i &    \chi|_{r\to\infty} &\to
  i \nonumber\\
  \label{space_bc}
  \partial_r\phi+\partial_r\bar\phi|_{r\to0} &\to 0  & \phi|_{r\to\infty} &\to i \\
  \phi-\bar\phi|_{r\to0}           &\to 0 \;.  && \nonumber
\end{align}
The conditions for the field $\phi$ at $r\to0$ make the original field
$\Phi$ regular at the origin.

\begin{figure}
  \begin{center}
    \includegraphics{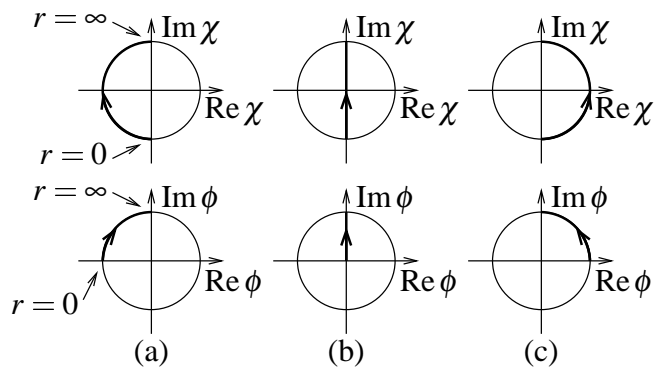}
  \end{center}
  \caption{Topological transition in the $SU(2)$ Higgs model: behavior
    of the fields $\phi$ and $\chi$.  Bold arrows show the change of
    the field as the radial coordinate increases from $r=0$ to
    $r=\infty$.  The configurations are shown: (a) at initial time,
    (b) in the middle of the process and (c) at final time.}
  \label{fig:transition}
\end{figure}

In this gauge no $r$-independent vacuum exists, but transition between
vacua with $n=0$ and $n=1$ is described in a very symmetric way.  The
behavior of the fields $\chi$ and $\phi$ for such transition is shown
in Fig.~\ref{fig:transition}.  In the original 4-dimensional model
this topology changing process corresponds to a transition where the
fields wind over the lower hemisphere of $SU(2)$ before the transition
and over the upper hemisphere after the transition.

The initial $\theta$--boundary conditions in gauge theory are quite
complicated.  The basic reason is that there is a redundant field
among the five fields $a_1$, $\phi$, $\bar\phi$, $\chi$, $\bar\chi$,
while the $\theta$-boundary conditions~\eqref{BVP_bc2} are to be
imposed on physical fields only.  The analytic expressions
for the modes $f_\bk$, $g_\bk$ in terms of the fields $\chi$, $\phi$,
$a_1$ are cumbersome (see Refs.~\cite{Rebbi:1996zx,Farhi:1996aq}) and
will not be presented here.  It is simpler, and more precise in the
lattice case, to perform this expansion numerically in the discretized
version of the model.  This expansion will be described in the
following section.

To complete the boundary value problem, one has to impose Gauss'
law and the equation fixing the time independent gauge invariance.  Note
that both of these equations are not full complex valued equations
(unlike the $\theta$--boundary conditions), otherwise the system would
have been overdetermined.  The point is that, the reality conditions
at final time~\eqref{BVP_real} forbid gauge transformations with
imaginary gauge functions and also guarantee that Gauss' law does
not have imaginary part.  So, only the real part of Gauss'
law~\eqref{gauss_law_eq} and equation fixing only real-valued gauge
transformations must be used.  Together with four $\theta$ boundary
conditions this gives the right number of boundary conditions for the
system with five complex valued fields $a_1$, $\alpha$, $\beta$,
$\mu$, $\nu$.  The exact form of the gauge fixing condition will be
given in section~\ref{sec:lattice}, because it is again most
conveniently expressed in lattice terms.

One more complication of the problem is the invariance of the
equations under translations along the real time.  To solve the
equations numerically this should be fixed in a controlled way, to
make sure the contour winds around the branching points of the
solution, and does not get too close to them.  A method of removing
this invariance will also be described in
Section~\ref{sec:lattice}.

\section{Computational challenges}
\label{sec:lattice}

\subsection{Discretized action}

To obtain a self-consistent system of equations, the discretization of
the equations~\eqref{a_eq}--\eqref{phibar_eq} should be done in a
gauge invariant way.

First, let us consider the discretized version of the action
\eqref{2dim_action}.  The spatial axis is discretized by introducing
sites $r_i$, $i=0,\dots,N$, where $r_0=0$, $r_N=L$.  The time grid
consists of sites $t_j$, $j=-1,\dots,N_t+1$.  We are working in the
$a_0=0$ gauge, and omit the subscript in the spatial component of the
gauge field, $a_1(r,t)\equiv a(r,t)$.  Field variables $\chi_{ij}$,
$\bar\chi_{ij}$ and $\phi_{ij}$, $\bar\phi_{ij}$ correspond to field
values on the space-time lattice sites, while $a_{ij}$ are defined on
spatial links and temporal sites.  We also absorb the $\Delta r_i$
factors in the definition of $a_{ij}$.  The boundary conditions in the
periodic instanton gauge, Eq.~\eqref{space_bc}, are
\begin{subequations}\label{lat_space_bc}
\begin{align}
  \chi_{0j}     &= -i &  \chi_{Nj}     &= i  \\
  \bar\chi_{0j} &= i  &  \bar\chi_{Nj} &= -i \\
  \phi_{0j} &= \half\left\{
                 \e^{-ia_{0j}/2}\phi_{1j}
                 +\e^{ia_{0j}/2}\bar\phi_{1j}
               \right\}
                  &  \phi_{Nj} &= i \\
  \bar\phi_{0j} &= \phi_{0j}
                  &  \bar\phi_{Nj} &= -i
\end{align}
\end{subequations}
for all $j$.  In the boundary condition for $\phi_{0j}$, the spatial
derivative in Eq.~\eqref{space_bc} was changed into a covariant one to
preserve exact lattice gauge invariance.  Thus the complex lattice field
variables left are
\begin{align*}
  \chi_{ij},\;
  \bar\chi_{ij},\;
  \phi_{ij},\;
  \bar\phi_{ij} && i&=1,\dots,N-1 & j&=-1,\dots,N_t+1 \\
  a_{ij}        && i&=0,\dots,N-1 & j&=-1,\dots,N_t+1\;,
\end{align*}
The discretized action reads
\begin{widetext}
\newsavebox{\eqboxa}\newsavebox{\eqboxb}\newsavebox{\eqboxc}
\sbox{\eqboxa}{$\begin{aligned}[t]
    \sum_{j=-1}^{N_t}\sum_{i=1}^{N-1}v_{1,j}\bigg\{ &
      w_{3,i}(\bar\chi_{i,j+1}-\bar\chi_{ij})(\chi_{i,j+1}-\chi_{ij})
     +w_{4,i}(\bar\phi_{i,j+1}-\bar\phi_{ij})(\phi_{i,j+1}-\phi_{ij})
     \bigg\}
  \end{aligned}$}
\sbox{\eqboxb}{$\begin{aligned}[t]
    -\!\sum_{j=-1}^{N_t}\sum_{i=0}^{N-1}v_{2,j}\bigg\{ &
      w_{2,i}(\e^{ia_{ij}}\bar\chi_{i+1,j}-\bar\chi_{ij})
             (\e^{-ia_{ij}}\chi_{i+1,j}-\chi_{ij})
 \\
    &
     +w_{1,i}(\e^{ia_{ij}/2}\bar\phi_{i+1,j}-\bar\phi_{ij})
             (\e^{-ia_{ij}/2}\phi_{i+1,j}-\phi_{ij})
     \bigg\}
  \end{aligned}$}
\sbox{\eqboxc}{$\begin{aligned}[t]
   -\!\sum_{j=-1}^{N_t}\sum_{i=1}^{N-1}v_{2,j}\bigg\{ &
     w_{5,i}\half(\bar\chi_{ij}\chi_{ij}-1)^2
   +w_{3,i}\left[
      \half(\bar\chi_{ij}\chi_{ij}+1)\bar\phi_{ij}\phi_{ij}
      +\frac{i}{2}\bar\chi_{ij}\phi^2_{ij}
      -\frac{i}{2}\chi_{ij}\bar\phi^2_{ij}\right]
 \\
   &
  +\lambda w_{4,i}(\bar\phi_{ij}\phi_{ij}-1)^2 \bigg\}\;,
  \end{aligned}$}
\begin{align}\label{disc_action}
  S              &=
    (S_\mathrm{ff}+S_\mathrm{t}+S_\mathrm{r}+S_\mathrm{int}) \\
  S_\mathrm{ff}  &=
    \sum_{j=-1}^{N_t}\sum_{i=0}^{N-1}v_{1,j}w_{1,i}
                                     (1-\cos(a_{i,j+1}-a_{ij}))
  \nonumber\\
  S_\mathrm{t}   &= \usebox\eqboxa
  \nonumber\\
  S_\mathrm{r}   &= \usebox\eqboxb \raisetag{11.5ex}
  \nonumber\\
  S_\mathrm{int} &= \usebox\eqboxc \raisetag{4ex}
  \nonumber
\end{align}
\end{widetext}
where the weights are
\begin{align*}
  v_{1,j} &= 1/\Delta t_{j+1/2} & v_{2,j} &= h_j\Delta t_j && \\
  w_{1,i} &= r_{i+1/2}^2/\Delta r_{i+1/2}
                                & w_{2,i} &= 1/\Delta r_{i+1/2}
    & w_{3,i} &= \Delta r_i \\
  w_{4,i} &= r_i^2\Delta r_i    & w_{5,i} &= \Delta r_i/r_i^2 &&
\end{align*}
with $\Delta t_{j+1/2}=t_{j+1}-t_j$, $\Delta t_j=(\Delta
t_{j+1}+\Delta t_j)/2$, and analogous expressions for $\Delta r_i$;
$h_j=1$ for $j=0,N_t$ and $1/2$ for $j=-1,N_t+1$.  Lattice field
equations are derived from~\eqref{disc_action} by variation over the
lattice fields, after the boundary conditions~\eqref{lat_space_bc}
have been used to exclude $\chi_{0j}$, $\bar\chi_{0j}$, $\phi_{0j}$,
$\bar\phi_{0j}$
and $\chi_{Nj}$, $\bar\chi_{Nj}$, $\phi_{Nj}$, $\bar\phi_{Nj}$
from the lattice action.

The action~\eqref{disc_action} is exactly invariant under
time-independent lattice gauge transformations of the form
\begin{eqnarray}
  a_{ij}    &\to& a_{ij}+\Omega_{i+1}-\Omega_{i}\;, \nonumber\\
  \chi_{ij} &\to& \e^{i\Omega_i}\chi_{ij}\;, \label{lattice_gauge}\\
  \phi_{ij} &\to& \e^{-\Omega_j/2}\phi_{ij}\;. \nonumber
\end{eqnarray}
This gauge freedom has to be fixed by boundary conditions.

\subsection{Boundary term: normal modes}

To obtain lattice version of \eqref{BVP_bc2} one notes that plane
waves are no longer eigenfunctions of the Hamiltonian on the lattice.
To find their analogue one brings the quadratic part of the action
\eqref{disc_action}, taken in the limit of continuous time, to the
canonical form.  We expand it near the space-independent vacuum
\eqref{triv_vac}
\begin{align}\label{bcgauge}
  \chi &= -i-\tilde{\chi} &  \bar\chi &= i-\tilde{\bar\chi} \\
  \phi &= -1+i\tilde\phi  &  \bar\phi &= -1-i\tilde{\bar\phi}
  \nonumber
\end{align}
(performing in the end a gauge transformation to the vacuum
\eqref{space_bc} is straightforward).
It is also useful to change to the notations~\eqref{defns},
\begin{align*}
  \tilde\chi       &= \tilde\alpha + i \tilde\beta &
  \tilde{\bar\chi} &= \tilde\alpha - i \tilde\beta \\
  \tilde\phi       &= \tilde\mu + i \tilde\nu &
  \tilde{\bar\phi} &= \tilde\mu - i \tilde\nu
\end{align*}
In these terms, the quadratic part of the action~\eqref{disc_action}
is
\begin{widetext}
\begin{align}
  S^{(2)} &= \frac{1}{2}\int dt
    \Biggl\{ \sum\limits_{i=0}^{N-1} w_{1,i}
    \dot a^2_i +  \sum\limits_{i=1}^{N-1} \left( 
    2w_{3,i} \dot{\tilde\alpha}_i^2 
    + 2w_{3,i} \dot{\tilde\beta}_i^2
    + 2w_{4,i}\dot{\tilde\mu}_i^2 
    + 2w_{4,i}\dot{\tilde\nu}_i^2 \right)\Biggr\}
  \nonumber\\
  & - \half\int dt \Biggl\{
    \sum\limits_{i=0}^{N-1} 
    2w_{2,i}\left[\left(a_i+\tilde\alpha_{i+1}-\tilde\alpha_i\right)^2 + 
    \left(\tilde\beta_{i+1}-\tilde\beta_i\right)^2\right]
  \nonumber\\
  &\hphantom{- \half\int dt \Biggl\{}
  +\sum\limits_{i=0}^{N-1} 
    2w_{1,i}\left[\left(a_i/2+\tilde\mu_{i+1}-\tilde\mu_i\right)^2
    + \left(\tilde\nu_{i+1}-\tilde\nu_i\right)^2\right]
  \nonumber\\
  &\hphantom{- \half\int dt \Biggl\{}
    + \sum\limits_{i=1}^{N-1} (4w_{5,i}+w_{3,i}) \tilde\beta_i^2 
    + \sum\limits_{i=1}^{N-1} w_{3,i} \left[
    \tilde\alpha_i^2 - 4 \tilde\alpha_i\tilde\mu_i + 4 \tilde\mu_i^2 \right]
    + 8\lambda \sum\limits_{i=1}^{N-1} w_{4,i}\tilde\nu_i^2 \Biggr\}\;.
  \label{quadratic-action}
\end{align}
\end{widetext}
As seen from Eq.~(\ref{quadratic-action}), the variables
$\tilde\alpha_0$, $\tilde\beta_0$, $\tilde\mu_0$ and $\tilde\nu_0$ do
not have kinetic terms.  Three of them are fixed by the boundary
conditions at $r=0$,
\[
  \tilde\alpha_0 = \tilde\beta_0 = \tilde\mu_0 =0.
\]
The fourth one, $\nu_0$, is determined from the field equation, which
for this variable reads
\[
  \tilde\nu_0=\tilde\nu_1.  
\]
After the variables with $i=0$ have been excluded in the manner
described above, the quadratic action takes the form
\[
  S = \int dt
    \left(\half d_I^2 \dot{\tilde\varphi}_I^2
      - \half \tilde\varphi_I S_{IJ} \tilde\varphi_J
    \right)\;,
\]
where the real valued coefficients $d_I$ and $S_{IJ}$ are to be read
off from Eq.~(\ref{quadratic-action}), indices $I,J$ label fields and
space points, and $\tilde\varphi_I$ stands for the fields
$\{\tilde\alpha_i,\tilde\beta_i,\tilde\mu_i,\tilde\nu_i,a_i\}$.  The
change of variables
\[
  y_I = d_I \tilde\varphi_I
\]
brings the kinetic term to the canonical form, 
\[
  S = \int dt\left(\half \dot y_I^2 - \half y_I \tilde S_{IJ} y_J \right)\;,
\]
where 
\[
  \tilde S_{IJ} = \frac{1}{ d_I} S_{IJ}
  \frac{1}{d_J} \;.
\]
The symmetric matrix $\tilde S_{IJ}$ is then diagonalized
\[
  \tilde S_{IJ} = O^T_{IK} \omega^2_K O_{KJ}, 
\]
where $O_{KJ}$ is an orthogonal matrix.  Introducing yet another set
of variables $z_I$ by the relations
\[
  O_{KJ} y_J = z_K; \hspace{1cm} y_J= O^T_{JK}z_K\;, 
\]
we finally bring the action to the diagonal canonical form, 
\[
  S =\int dt\left( \half \dot z_I^2 - \half \omega^2_I z_I^2\right)\;.
\]
Therefore, vectors 
\[
  \xi_{(K)I} = O_{KI}
\]
are normal modes in the lattice formulation of the theory, and should
be used instead of the usual spherical waves.  The corresponding
frequencies are $\omega^2_K$.

The matrix $O_{KJ}$ and frequencies $\omega_K$ are found numerically.
Since they depend only on the spatial lattice parameters (size and
spacing) and coupling constant $\lambda$, and do not depend on the
background vacuum field configuration, it is sufficient to perform
this diagonalization once for a given lattice.  The first $4N-3$
eigenvectors $\xi_{(K)I}$ and eigenvalues $\omega_K$ correspond to
physical modes, and the rest $N-1$ of them have $\omega=0$ and thus correspond
to the gauge (unphysical) degrees of freedom.

\subsection{Boundary conditions.}
\label{sec:lattice_bc}

\paragraph{$\theta$ boundary conditions.}
To derive the lattice version of the boundary conditions, one takes
the variation of the exponent for the total probability
\eqref{smallPIsigmaEN}, which can be written in the following form
\begin{multline}\raisetag{9ex}\label{W}
  iS(z)-iS(z') \\
  -\half \frac{\omega}{ 1-\gamma^2} 
  \left\{ (1+\gamma^2) (z_{-1}^2+{z'}_{-1}^2) - 4\gamma z_{-1}z_{-1}'\right\} + \dots
\end{multline}
where dots denote terms irrelevant in the current context and
$\gamma=\e^{-\theta}$.  One has to vary the lattice version of (\ref{W})
with respect to $z_{I,-1}$ (values of $z$ at the first time slice) and set
$z'=z^*$.  The variational equation reads
\[
  i \frac{\delta S}{ \delta z_{I,-1}} 
  - \frac{\omega_I}{ 1-\gamma^2} (1+\gamma^2) z_{I,-1}
  + \frac{2\omega_I\gamma}{ 1-\gamma^2} z_{I,-1}^*=0
\]
which leads to
\begin{equation}\label{num-bc}
  \frac{\delta S}{ \delta z_{I,-1}} + i \frac{1-\gamma}{ 1+\gamma} \omega_I
  \Real z_{I,-1} - \frac{1+\gamma}{ 1-\gamma} \omega_I \Imag z_{I,-1}=0.
\end{equation}
where the derivatives of the action are equal to the classical momenta
of the modes
\begin{equation*}
  \frac{\delta S}{ \delta z_{I,-1}} = -v_{1,-1}(z_{I,0}-z_{I,-1})\;.
\end{equation*}
Here the index $I=1,\ldots,4N-3$ labels physical degrees of freedom.  One can
go back to the original notations by means of the relation
\[
  z_I = \xi_{IJ} d_J \tilde\varphi_J,
\]
where $I=1,\ldots,4N-3$.

Finally, we use the gauge transformation
\[
  \Omega_i=\pi\exp\left(-\frac{r_i}{c(L-r_i)}\right)
\]
with $c=0.5$ to transform the fields from the periodic instanton
gauge~\eqref{space_bc} to the form $\tilde\varphi_J$ \eqref{bcgauge}
(of course, other gauge choices with $\Omega(0)=\pi$, $\Omega(L)=0$
are possible).  This transformation has the form
\begin{equation}\label{eq:gg1}
  \tilde\varphi_J=g_{KJ}\varphi_J-\tilde\varphi^\mathrm{vac}_J\;,
\end{equation}
where $\varphi_J$ stands for the fields
${\alpha_j,\beta_j,\mu_j,\nu_j,a_j}$ in periodic instanton gauge.  The
matrix $g_{KJ}$ and vector $\tilde\varphi^\mathrm{vac}_J$ can be
easily read off from the expression for the lattice gauge
transformation~\eqref{lattice_gauge} and definition~\eqref{bcgauge}.

\paragraph{Zero modes part.}

The $\theta$-boundary conditions~\eqref{num-bc} give only $4N-3$
(complex) equations, while $5N-4$ boundary conditions are required at
the initial time.  The ``left over'' $N-1$ conditions correspond to
$N-1$ gauge degrees of freedom in the model.  As described in
section~\ref{sec:spher}, for these $N-1$ equations one has to use the
real part of Gauss' law~\eqref{gauss_law_eq},
\begin{multline}\label{num-gauss}
  \Real\Big[w_{1,i-1} \sin (a_{i-1,-1}-a_{i-1,0})-w_{1,i} \sin
    (a_{i,-1}-a_{i,0})
  \\
  + i w_{3,i} (\bar\chi_{i,-1}\chi_{i,0}-\chi_{i,-1}\bar\chi_{i,0})
  \\
  + \frac{i}{2} w_{4,i} 
  (\bar\phi_{i,-1}\phi_{i,0}-\phi_{i,-1}\bar\phi_{i,0}) \Big] = 0\;,
\end{multline}
where $i=1,\ldots,N-1$.  One also makes use of equations that fix the
remaining real gauge freedom.  The latter equations are
\begin{equation}\label{num-gauge}
  \Real z_{L,-1} = 0
\end{equation}
for all $L=4N-2,\ldots,5N-4$.  These modes have zero frequency
$\omega_L=0$ and correspond to the unphysical degrees of freedom which
change under gauge transformation, so Eq.~\eqref{num-gauge} fixes the
residual gauge invariance with real gauge functions completely.  Gauge
transformations with imaginary gauge functions are forbidden by the
reality conditions at final time.

\paragraph{Final boundary conditions.}

It is straightforward to implement the reality conditions at
final time \eqref{BVP_real}.  Supposing that the last two time grid
points, $N_t$ and $N_t+1$ are on the real time axis, they are
\begin{align}
  &\Imag\alpha_{i,N_t} = \Imag\beta_{i,N_t} = 0
  \nonumber\\
  &\Imag\mu_{i,N_t} = \Imag\nu_{i,N_t} = 
  \Imag a_{i,N_t} = 0
  \label{lattice_realbc}\\
  &\Imag\alpha_{i,N_t+1} = \Imag\beta_{i,N_t+1} = 0
  \nonumber\\
  &\Imag\mu_{i,N_t+1} = \Imag\nu_{i,N_t+1} = 
  \Imag a_{i,N_t+1} = 0\;.
  \nonumber
\end{align}
For energies below the bifurcation line $E_1(N)$ the time $N_t$
can be chosen to coincide with the point C of the time contour (so
there are only two lattice points $N_t$ and $N_t+1$ on the whole CD
part).  For higher energies, though, the fields are not real along the
most part of the real axis, so the part CD of the time contour has to
be as long as possible (see Section~\ref{sec:reg}).

\paragraph{Fixing time translational invariance.}
One more complication is that, in the continuous formulation, the boundary
value problem~(\ref{final_BVP}) has an invariance under translations
along real time (both field equations and boundary conditions are
invariant under such a translation).  To define properly the boundary
value problem, one has to fix the position of the solution in time.
In the lattice version this invariance is violated by the
discretization and finite volume effects, but this violation does not
enable one to control the position of the time contour relative to the
branching points of the solution.

The existence of this invariance means that one of the equations is
redundant (if discretization and finite volume effects are discarded).
Somewhat arbitrarily, we take as redundant one of the real equations
entering the $\theta$-boundary conditions~\eqref{BVP_bc2}
\begin{equation}\label{arg_thetabc}
  \arg f_\bk = \arg g_\bk \;,
\end{equation}
for a specific mode.  Provided the system linearizes at initial time,
this equation is indeed a consequence of the others.  The reason is that
reality conditions at final time imply that the (conserved) energy is
real.  Hence the (linearized) energy~\eqref{BVP_E} is real at initial
time.  Then one of the modes automatically obeys
Eq.~\eqref{arg_thetabc} provided all other modes obey the
$\theta$-boundary condition~\eqref{BVP_bc2}.

This suggests the following modification of the equations.  One of the
equations \eqref{BVP_bc2} is changed to
\[
  |\fk| = \e^{-\theta} |g_{\bk}| \;,
\]
whose lattice version is (cf.\ Eq.~\eqref{num-bc})
\begin{widetext}
\begin{equation}\label{num-fixtheta}
  \left(1-\gamma^2\right)\left[
    \left|\frac{\delta S}{\delta z_{K,-1}}\right|^2+
    \omega_K^2\left|z_{K,-1}\right|^2
  \right]
  -2\omega_K\left(1+\gamma^2\right)\left[
    \Real\frac{\delta S}{\delta z_{K,-1}}\Imag z_{K,-1}
    -\Imag\frac{\delta S}{\delta z_{K,-1}}\Real z_{K,-1}
  \right]
  =0\;.
\end{equation}
\end{widetext}
Thus, instead of Eq.~\eqref{arg_thetabc} one imposes another boundary
condition, which is not invariant under time translations.  The choice
of the latter is a matter of convenience.  We control the position of
the solution in time by imposing the boundary condition that fixes the
``center-of-mass'' of the field $\chi$ at the initial time to be equal to 
a given $R$:
\begin{equation}\label{num-fix}
  \Real\sum_{i=1}^{N-1}
  w_{4,i}(r_i-R)(\chi_{i,-1}\bar\chi_{i,-1}-1)^2=0\;.
\end{equation}
This prescription works if the mode $z_K$ in
Eq.~\eqref{num-fixtheta} is reasonably occupied at the initial time,
otherwise the equation~\eqref{arg_thetabc}, which is ``thrown away'',
is nearly degenerate.  Aside from this, the results of the
calculations do not depend significantly on the mode chosen.

The relative phase between $f_\bk$ and $g_\bk$ can be used to check
the validity of the calculations.  In the linear regime it should be
equal to zero, so the actual value of this phase indicates how close
the system is to the linear regime at the initial time.

To summarize, the lattice boundary value problem consists of the field
equations, obtained from action~\eqref{disc_action} for all inner
lattice points ($i=1,\ldots,N-1$, $j=0,\ldots,N_t$, a total of
$(N-1)(N_t+1)$ equations), the final reality boundary
conditions~\eqref{lattice_realbc} ($N-1$ equations), the $\theta$ boundary
conditions~\eqref{num-bc} for all modes except one mode $z_K$ ($N-2$
equations), and a pair of real equations~\eqref{num-fixtheta},
\eqref{num-fix} (one complex valued equation).  This makes
$(N-1)(N_t+3)$ complex equations for the same number of variables.

\subsection{Search for solutions}
\label{sec:solution_search}

The equations to be solved make a set of discretized partial
differential equations which change their signature from hyperbolic on
the Minkowskian parts of the time contour to elliptic on the Euclidean
part.  The problem at hand is a boundary value problem which cannot be
transformed into an initial value one.  This means that the equations
can be solved only globally, as a set of nonlinear equations at all
$r,t$ grid coordinates.

To deal with the non-linear system of equations we employ a
multidimensional analog of the Newton--Raphson method which approaches
the desired solution iteratively.  At each iteration, the
\emph{linearized} equations in the background of the current
approximation are solved.  The next approximation is obtained by
adding the solution to the background, and the procedure is repeated.
The advantage of the algorithm is that it does not require
positive-definiteness of the matrix of second derivatives.  It is,
however, sensitive to zero modes. In the absence of zero modes, the
algorithm converges quadratically; the accuracy of $10^{-9}$ is
typically reached in 3-5 iterations. The convergence slows down in the
presence of very soft modes, as typically happens near bifurcation
points.

\subsection{Elimination algorithm}
\label{sec:elimination}

The discrete version of the equations derived from \eqref{disc_action} is
\[
  \frac{\partial S}{\partial\varphi_{jI}} = 0
\]
(here $\varphi_{jI}=\{a,\alpha,\beta,\mu,\nu\}(t_j,r_i)$ and $I$ runs
from $0$ to $5N-4$, $j=0\dots N_t$).  The Newton--Raphson iteration is
\begin{multline}\label{disceq}
  \frac{\partial^2 S}{\partial\varphi_{jI}\partial\varphi_{j-1,K}}
    u_{j-1,K}
  +\frac{\partial^2 S}{\partial\varphi_{jI}\partial\varphi_{jK}}
    u_{jK} \\
  +\frac{\partial^2 S}{\partial\varphi_{jI}\partial\varphi_{j+1,K}}
    u_{j+1,K}
  +\frac{\partial S}{\partial\varphi_{jI}} = 0
\end{multline}
(all other second derivatives are zero) and before the next
Newton--Raphson step the fields are changed according to
\[
  \varphi^{(n+1)}_{jI} = \varphi^{(n)}_{jI}+u_{jI}
\]
Equations~\eqref{disceq} can be rewritten in matrix form
\begin{equation}\label{tridiag1}
  \tilde D^{(-)}_j\cdot u_{j-1} +\tilde D_j\cdot u_j
  +\tilde D^{(+)}_j\cdot u_{j+1}+\tilde b_j = 0
\end{equation}
where $u_j$ and
$b_j=\frac{\partial S}{\partial\varphi_{jI}}$
are $(5N-4)$--dimensional vectors,
$
 \tilde D^{(-)}_j =
 \frac{\partial^2 S}{\partial\varphi_{jI}\partial\varphi_{j-1,K}}
$,
$
 \tilde D_j =
 \frac{\partial^2 S}{\partial\varphi_{jI}\partial\varphi_{jK}}
$,
$
 \tilde D^{(+)}_j =
 \frac{\partial^2 S}{\partial\varphi_{jI}\partial\varphi_{j+1,K}}
$
are $(5N-4)\times(5N-4)$ matrices.  By multiplying Eq.~\eqref{tridiag1}
by $\tilde{D}^{-1}_j$ we get
\begin{equation}\label{tridiag}
  u_j = D^{(-)}_j\cdot u_{j-1}+D^{(+)}_j\cdot u_{j+1}+b_j
\end{equation}
with $D^{(\pm)}_j=-\tilde{D}^{-1}_j\cdot\tilde{D}^{(\pm)}_j$,
$b_j=-\tilde{D}_j^{-1}\cdot\tilde{b}_j$.  This system of linear
equations was solved by the following version of
``divide--and--conquer'' elimination algorithm.  Excluding $u_j$ for
some $j$ gives
\begin{multline*}
  u_{j-1} = \left(1-D^{(+)}_{j-1}\cdot D^{(-)}_j\right)^{-1}\times \\
     \Bigl[
      D^{(-)}_{j-1}\cdot u_{j-2} 
      +D^{(+)}_{j-1}\cdot D^{(+)}_j\cdot u_{j+1} \\
      +(D^{(+)}_{j-1}\cdot b_j+b_{j-1})
    \Bigr]
\end{multline*}
\begin{multline*}
  u_{j+1} = \left(1-D^{(-)}_{j+1}\cdot D^{(+)}_j\right)^{-1}\times \\
     \Big[
      D^{(-)}_{j+1}\cdot D^{(-)}_j\cdot u_{j-1}
      +D^{(+)}_{j+1}\cdot u_{j+2} \\
      +(D^{(-)}_{j+1}\cdot b_j+b_{j+1})
    \Big]
\end{multline*}
Since the elimination of an equation changes only adjacent equations,
it is possible to eliminate all equations with odd $j$ in parallel,
and arrive to a system of the type~\eqref{tridiag} again, but with two
times less variables and equations.  This is the second level of
elimination.  After a series of eliminations we arrive at a system of
only two equations for $j=0$ and $j=N_t$:
\begin{subequations}\label{after_elim}
\begin{align}
  u_0 &=
    \hat{D}^{(-)}_0\cdot u_{-1}+\hat{D}^{(+)}_0\cdot u_{N_t}+\hat{b}_0
  \\
  u_{N_t} &=
    \hat{D}^{(-)}_{N_t}\cdot u_{0}+\hat{D}^{(+)}_{N_t}\cdot u_{N_t+1}
    +\hat{b}_{N_t}
\end{align}
\end{subequations}
where $\hat{D}^{(\pm)}$ and $\hat{b}$ have the values resulting from
the elimination of all intermediate equations.  Solving them together
with the boundary conditions%
\footnote{Unlike the field equations at the intermediate
points~\eqref{disceq}, which are analytic, the boundary conditions
involve complex conjugation.  So, equations~\eqref{after_elim} and
boundary conditions are to be viewed as eight real matrix equations.  All the
elimination calculations (and reconstruction of field values
afterwards) can be done, however, with complex algebra, which is two
times more efficient.}
(also linearized), which involve $u_{-1}$ and $u_0$ for initial
boundary condition and $u_{N_t}$ and $u_{N_t+1}$ for the final one, we
determine the corrections $u_{-1}$, $u_0$, $u_{N_t}$ and $u_{N_t+1}$.
Then it is straightforward to reconstruct $u$ at all intermediate
points, using the equations~\eqref{tridiag} for each elimination
level.

\subsection{Solutions below the sphaleron energy}
\label{sec:below_sph}

The Newton--Raphson method requires a good initial approximation for
the solution.  This favors the following general strategy.  We first
find the periodic instanton solution, which corresponds to
$\theta=0$~\cite{Bonini:2000mb} and can be obtained via a minimization
procedure.  After the periodic instanton is found, we change the
parameters $T$ and $\theta$ in small steps, using the solution from
the previous step as a starting configuration.  At each step we then
calculate the energy $E$, number of particles $N$ and the suppression
exponent $F(E,N)$ for the solution obtained.

\begin{figure*}
  \begin{center}
\includegraphics{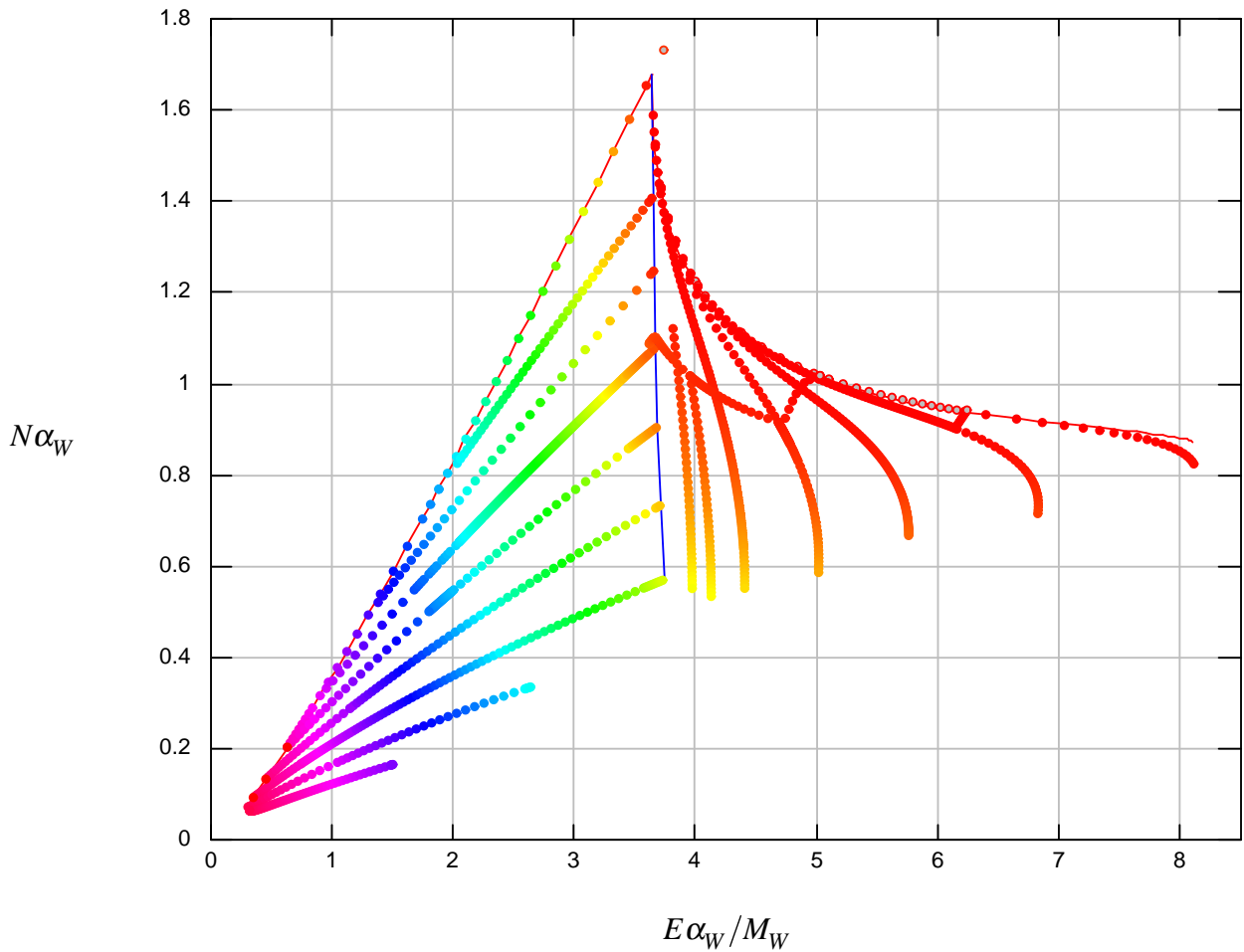}
  \end{center}
  \caption{Search for solutions.  Each point corresponds to one
    solution of the boundary value problem.  The color of the points
    tracks the suppression exponent $F(E,N)$.  The almost vertical line is
    the line of bifurcations $E_1(N)$, (cf.\ Fig.~\ref{fig:regions}).}
  \label{fig:search}
\end{figure*}

This procedure is illustrated in Fig.~\ref{fig:search}, where each dot
represents one solution of the boundary value problem.  Initial
periodic instanton configurations correspond to the points on the
upper left line in the figure.  Starting from these points, the value
of $\theta$ was increased, and lines with constant values of $T$ were
obtained until the bifurcation line was met.  Data obtained in this
way make almost straight lines in the left part of
Fig.~\ref{fig:search}.

The boundary value problem~\eqref{final_BVP}--\eqref{BVP_N} does not
explicitly refer to the topological properties.  Hence, it is not
guaranteed that its every solution describes a transition between
topologically distinct vacua.  This is not a problem at $\theta=0$,
because of the proper topological structure of the periodic instanton
solutions.  But at non-zero $\theta$ one should check that the
solution indeed has correct topology.

\begin{figure}
  \begin{center}
    \includegraphics{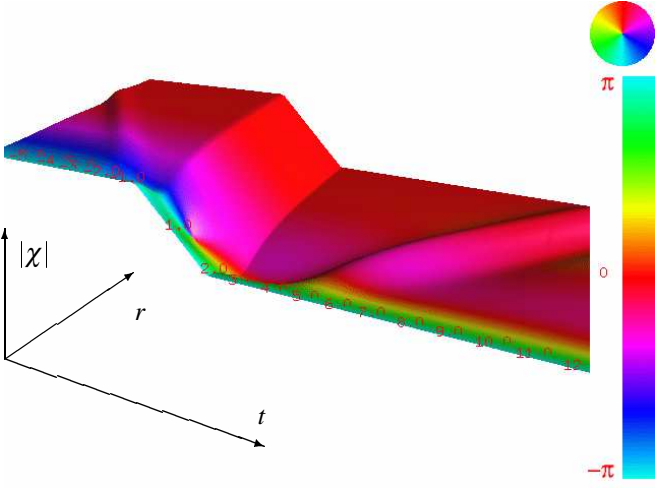}
  \end{center}
  \caption{Visualization of the field $\chi$ for a
    solution with $N=1$ and $E=3.35$.  The color tracks the phase of the
    field.  The part corresponding to the Euclidean evolution is
    inclined for visualization purposes.}
  \label{fig:surface_good}
\end{figure}

The topological properties of a given solution are associated with the
behavior of the phases of the fields, see Fig.~\ref{fig:transition}.
A very useful tool to control the properties of the solution is
visualization of the field behavior.  The visualization of a
representative field configuration is presented in
Fig.~\ref{fig:surface_good}.  It describes the field $\chi(r,t)$, with
the phase of the field encoded in color.  The Euclidean part of the
time contour is inclined to make it distinct from the Minkowskian
parts.  In the initial state (left part of the surface) the field is
close to its vacuum value, with excitation in the form of the incoming
spherical wave moving towards to $r=0$.  The final state (right part
of the surface) contains the outgoing wave.  The phase of the field
clearly behaves differently in the initial and final states.  This
confirms that the topological transition indeed has occurred (compare
to upper three images in Fig.~\ref{fig:transition}).  Several other
important properties of the solution may also be seen immediately.
These are: the moment when the field goes through $\chi=0$ in the
middle of the Euclidean evolution, which of course should happen with
the field evolving between neighboring vacua; the wide outgoing wave,
suggesting that a large number of low energy particles is created after
the transition; the small and relatively sharp incoming wave, meaning that
higher energy modes are occupied and the number of particle in the
incoming state is smaller.

\section{Going over the sphaleron energy}
\label{sec:reg}

\begin{figure}
  \begin{center}
    \includegraphics{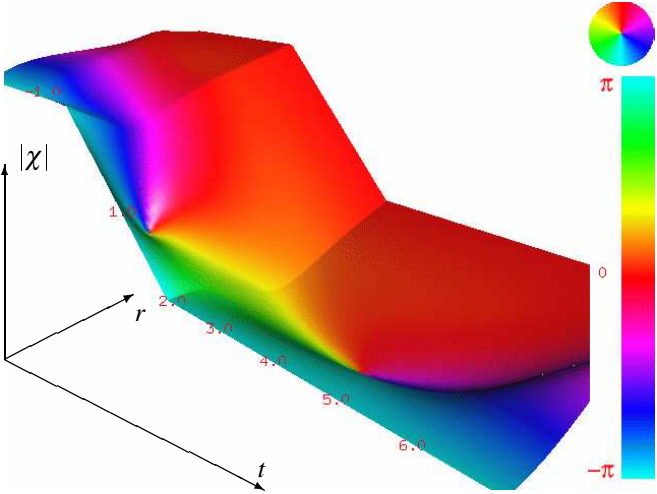}%
    \caption{Solution for $T/2=2$ and $\theta=3.35$, without
      regularization.  For this solution $E/E_\mathrm{sph}=1.04$,
      $N\alpha_W=0.94$, so that $E>E_1(N)$.  One observes that the
      topological properties of the solution are wrong: it begins and
      ends in the same vacuum.}
    \label{fig:bad_topology}
  \end{center}
\end{figure}

The procedure described above works as it is for relatively low
energies $E\lesssim E_\mathrm{sph}$ only.  With growing energy, the
solutions on the CD part of the contour tend to stay for a long time
close to the sphaleron.  As the energy approaches some $N$-dependent value
$E_1(N)$ this time tends to infinity, and if one continues to search
for solutions to the boundary value problem~\eqref{final_BVP} with
reality condition imposed at finite positive time, the solutions above
this energy have wrong topological properties, i.e.\ they end up in
the same topological vacuum as the initial one (see
Fig.~\ref{fig:bad_topology}).  This situation is not specific to the
$SU(2)$ gauge model studied here, but appears quite generally in quantum
mechanical tunneling with multiple degrees of freedom.  It was
observed also in the study of the false vacuum decay in scalar field
theory ~\cite{Kuznetsov:1997az} and in quantum
mechanics with two degrees of 
freedom~\cite{Bonini:1999cn,Bonini:1999kj}.  The phenomenon was
studied in detail in Ref.~\cite{Bezrukov:2003yf} in the case of
quantum mechanics of two degree of freedom, and a general method of
dealing with this difficulty was proposed there and checked against
the exact solution of the Schr\"odinger equation.  We describe here
its gauge field version.

As suggested in Ref.~\cite{Bezrukov:2003yf}, the line $E_1(N)$ is the
bifurcation line at which two types of solutions to the boundary value
problem~\eqref{final_BVP} meet.  These are (i) solutions which end up
close to the same vacuum as the initial one and (ii) solutions that
arrive at the sphaleron with excited positive modes (in the case of
field theory these excitations fly away quickly in the form of
spherical waves in the sphaleron background).  The former solutions
are unphysical, while solutions of the latter form determine the
tunneling exponent.  For the interesting solutions of type (ii), the
condition~\eqref{BVP_real} is satisfied only asymptotically, so it is
very hard to find them numerically.  A way out is to introduce a small
regularization parameter into the equations of motion, which would not
allow a solution to stay close to the sphaleron for infinite time.
The final result is then obtained in the limit of zero regularization
parameter.

To implement these ideas we start with the regularized expression for
the cross section
\begin{equation}\label{reg_sigmaEN}
  \sigma_\epsilon(E,N) =
    \sum_{i,f} |\bra{f}
      \e^{-2\epsilon \hat{T}_\mathrm{int}}
      \hat{S}
      \hat{P}_E \hat{P}_N
    \ket{i}|^2 \;,
\end{equation}
where $\epsilon$ is a small parameter and $T_\mathrm{int}$ is a
functional proportional to the time the system spends in the
interaction region.  In case of gauge--Higgs theory we use the
functional
\begin{equation}\label{Tint}
  T_\mathrm{int} = \int\! dt \int\! dr\, (\bar\phi(r)\phi(r)-1)^8 \;.
\end{equation}
The path integral for~\eqref{reg_sigmaEN} is no longer saturated by
classical solutions spending infinite time close to the sphaleron,
while the original cross section $\sigma(E,N)$ is obtained in the
limit $\epsilon\to0$.  The boundary value problem
for~\eqref{reg_sigmaEN} coincides with the unregularized one, but with
the action modified by adding an imaginary term of the form
\begin{equation}\label{reg-dS}
  \delta S = i\epsilon\int\! dt \int\! dr\, (\bar\phi(r)\phi(r)-1)^8 \;.
\end{equation}
The equations of motion \eqref{2dim_eqs} are modified accordingly.

The functional~\eqref{Tint} has several important features: i) it is
gauge invariant; ii) it is large and positive on configurations close
to the sphaleron (where the fields stay for a long time away from their
vacuum values); iii) it does not change the free dynamics in the
linear region, since it does not produce quadratic terms in the
expansion of the action about vacuum (this is important for the
boundary conditions to be unaffected).

With this regularization one obtains results for all energies $E$.
The procedure is as follows.  One introduces small but non-zero
$\epsilon$ at energies below $E_1(N)$, then obtains solutions with
proper topology for any energy.  Then one takes the limit
$\epsilon\to0$.  Upon taking this limit, the configurations with
$E>E_1(N)$ stay for longer time close to the sphaleron, which means
that in the limit of $\epsilon\to0$ the solution tunnels ``onto'' the
sphaleron.

Moreover, at the boundary of the classically allowed region, the
solutions to the regularized problem merge smoothly with the classical
over-barrier topology changing solutions, because the bifurcation on
the boundary of the classically allowed domain is regularized exactly
in the same way as the bifurcation at $E=E_1(N)$.  At the boundary of
the classically allowed region $F=0$ by definition, so the regularized
version of this functional $F_\epsilon$ is proportional to $\epsilon$.
This means that $T$ and $\theta$ are also proportional to $\epsilon$
there, and as the regularization is turned off, $\epsilon\to0$, both
$T$ and $\theta$ disappear, leading to purely real classical boundary
value problem in real time.

There is one more complication in the $SU(2)$ field theory, which is
relevant to this procedure.  For energies $E<E_1(N)$ the amount of
time $T$ spent on the Euclidean part of the contour is a growing
function of energy (as opposed to the situation in two dimensional
quantum mechanics~\cite{Bezrukov:2003yf}), while it is zero at the
boundary of the classically allowed region $E_0(N)$.  This means that
$T$ as function of energy at fixed $\theta$ has a maximum somewhere in
between $E_1(N)$ and $E_0(N)$ (the calculations show that the maximum
is actually at $E_1(N)$, see Fig.~\ref{fig:ne_t}).  For $T$ close to
this maximum value, the Newton-Raphson method fails because of the
presence of two nearby solutions with equal values of $T$.  This new
bifurcation is absent, if one searches for solutions 
with \emph{fixed energy $E$,}
instead of fixed $T$. To formulate the boundary
value problem with fixed $E$ instead of fixed $T$, one simply sets
the contour height $T'$ to some conveniently chosen value, and leaves
$T$ as a free variable.  This leads to a trivial modification of the
initial boundary conditions,
\begin{equation}\label{ttprime-bc}
  \fk = \e^{-\theta-\omega\bkind(T-T')} g\bkind\;.
\end{equation}
An additional equation is then required to find one extra undetermined
variable $T$.  This is the equation involving energy of the solution,
Eq.~\eqref{BVP_E}.  With this modification of the procedure, the
``bifurcation'' corresponding to the maximum of $T$ disappears.  Note
however, that this method had to be applied with great care---on
realistic grids it is hard to achieve fully linear regime in the
initial state, therefore the difference between $T$ and $T'$ must be
small, as the dependence of the fields on imaginary time is
exponential.  In our calculations, the version of the boundary value
problem with fixed $E$ ($T'$ different from $T$) was used only to
cross the maximum of $T$.

In our numerical calculations we introduced small non-zero $\epsilon$
as the energy of the solutions approached $E_1(N)$.  Simultaneously,
the modification~\eqref{ttprime-bc} was used to get past the maximum
of $T$ at $E_1(N)$.  In Fig.~\ref{fig:search} the solutions obtained
with this modification are represented by points on the line that
crosses the bifurcation line $E_1(N)$.  At higher energies the
modification~\eqref{ttprime-bc} is no longer needed, and only
regularization with non-zero $\epsilon$ was used.  In
Fig.~\ref{fig:search} solutions to the regularized problem in region
A.II correspond to points on the curved lines in the right part of the
plot (lines of constant $T$).  The line with the highest energy has
zero suppression exponent and corresponds to the boundary of the
classically allowed region.  To connect lines of constant $T$, we
obtained a set of solutions represented by the irregular line in right
part of Fig.~\ref{fig:search}.

\section{Numerical results}
\label{sec:num_results}

There are several factors affecting the choice of the lattice size and
shape.  The physical spatial size of the lattice $L$ is chosen large
enough to make comfortable room for the sphaleron.  More importantly,
$L$ determines how close to the linear regime the system is in the
initial state: the farther away from the origin, the smaller
becomes the amplitude of incoming spherical
waves.  After $L$ is chosen, the lengths of
the parts AB and CD of the time contour are determined completely;
the length of the AB part $T_{AB}$ is slightly smaller than $L$, so that
the incoming wave does not reach the spatial boundary $r=L$.  The
length of the CD part $T_{CD}$ is zero for energies below the
bifurcation energy $E<E_1(N)$.  For higher energies, $T_{CD}$ is
adjusted to be long enough, so that the solution gets close to the
vacuum configuration, and the regularization~\eqref{reg-dS} does not
contribute significantly to the equations of motion at the final
moment of time when the reality boundary conditions are imposed.

The lattice spacing $\Delta r$ constrains the precision of the
discretization in two different ways.  First, it is chosen to be substantially
smaller than the size of the instanton-like part of the configuration,
i.e.\ the characteristic scale of nonlinear dynamics that occurs near
$r=0$ during the topological transition itself.  Second, $\Delta r$
controls the energy of the hardest mode in the initial state, thus
limiting the lowest particle number $N$ that can be reached for given
energy $E$.  The time spacing $\Delta t$ is chosen to be smaller than
$\Delta r$ to guarantee stability of the numerical procedure.

The amount of computer memory required for a lattice of spatial size $N_r$
and time length $N_t$ is approximately $2\times N_t(5N_r)^2\times16$
bytes (see Section~\ref{sec:elimination}), while the CPU time of one
Newton--Raphson iteration scales roughly as $N_t(5N_r)^3$.  It was
noted in Section~\ref{sec:elimination} that the algorithm is suitable
for parallel execution, so one divides this time by the number of
processors available for the calculations\footnote{The parallelization
algorithm is effective only if $N_\mathrm{processors}<\sqrt{N_t}$, so
the shortest possible wall clock time in an ideal situation is proportional
to $\sqrt{N_t}(5N_r)^3$}.  Overall, the most strongly constrained is the
spatial size $N_r$: a two times larger spatial grid means eight times
longer processor time.

The main results we present in this paper were obtained on a grid with
spatial size $L=8$ (i.e.~$L=8/\sqrt{2}M_W$) and number of spatial grid
points $N_r=90$.  The length of the initial Minkowskian part of the
contour $T_{AB}$ was equal to 6.  The number of time grid points $N_t$ on
the part AB of the contour was equal to 200, while on the Euclidean
part BC it was equal to 150.  The number of points on CD part varied
from 2 for energies $E<E_1(N)$ to about 400 for higher energies
(when the $\epsilon$--regularization was used).  On the largest grids
the amount of memory used was 4Gb, and it took 3 minutes for one
Newton--Raphson iteration on a 16 processor IBM-RS/6000 supercomputer,
or about 15 minutes for one full solution.

We obtained the results for the suppression factor in the region of
$E$ and $N$ shown in Fig.~\ref{fig:search}.  For the lattice parameters
we used, this region is limited mainly by the effects of nonlinearity at
the initial time, preventing us from reaching smaller particle numbers.
When energy and particle number are small simultaneously (bottom-left
part of the plot), effects of the spatial discretization (finite $\Delta r$)
are also important.

To check the discretization effects, a limited set of calculations was
performed on smaller grids.  The results presented here coincide with
results obtained with $N_r=64$ with precision better 
than 1\% (except for very small
energies).  With $N_r=45$, on the other hand, the results coincide only
for sufficiently large initial particle numbers, exactly as one would expect.

\begin{figure}
  \begin{center}
    \includegraphics[width=\columnwidth]{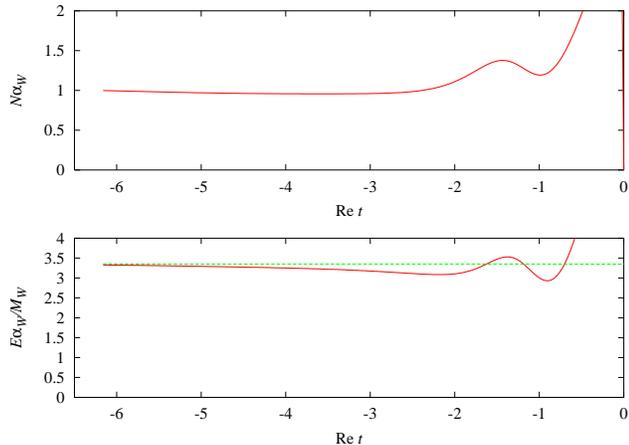}
    \caption{Particle number evolution (upper plot) and linear energy
      evolution (lower plot) for the configuration with $N=1$ and
      $E=3.35$.  The exact (full nonlinear) energy is also plotted for
      reference (straight dotted line).}
    \label{fig:linenergy}
  \end{center}
\end{figure}

The linearization of the system in the initial state can be checked by
evaluating the time dependence of the linear energy~\eqref{BVP_E} and
particle number~\eqref{BVP_N} on the part AB of the contour.  For
linearized system, these should be independent of time.  For a typical
configuration this test is shown in Fig.~\ref{fig:linenergy}.  The
linear energy coincides with the exact one in the initial state with
precision of order of 1\% or better, which confirms that the
solution is quite close to the linear regime.  Another test of
linearity is the amount of the violation of the initial boundary
condition~\eqref{arg_thetabc}, which is discarded to impose the time
translation invariance fixing relation (Section~\ref{sec:lattice}).
This amount grows towards smaller $N$, and apparently this is one of
the effects preventing us from going to lower $N$ with the current
spatial lattice size $L=8$.  Larger lattices are needed to achieve
better linearization on the initial part of the time contour and thus
reach smaller particle numbers.

We made additional checks of the precision of the numerical
calculations, including conservation of energy
and the inverse Legendre transform
\begin{eqnarray} 
  \theta &=& -(4\pi)\left.\frac{\partial F}{\partial N}\right|_E
  \label{dfdn} \\
  T      &=& -(4\pi)\left.\frac{\partial F}{\partial E}\right|_N  \;.
  \label{dfde}
\end{eqnarray}
These checks are satisfied with precision better than $10^{-3}$.  This
means that the precision of the final results is determined mostly by the
quality of the linearization in the initial state
(of the order of 1\%).

\begin{figure}
  \centerline{\includegraphics[width=\columnwidth]{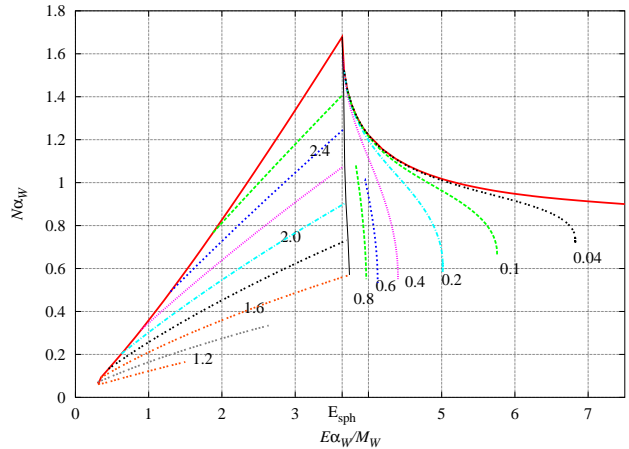}}
  \caption{Lines of constant $T$.}
  \label{fig:ne_t}
\end{figure}

\begin{figure}
  \centerline{\includegraphics[width=\columnwidth]{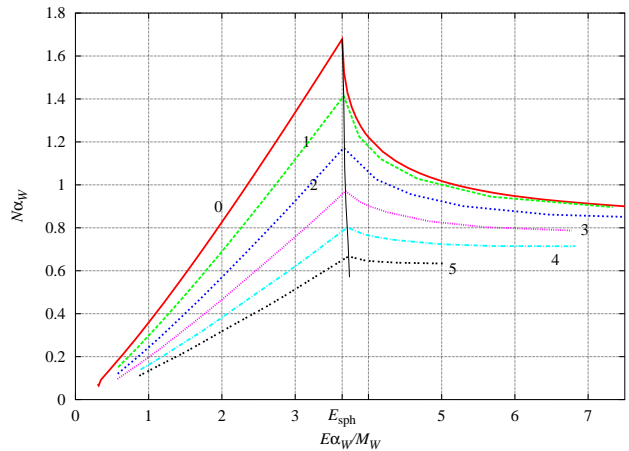}}
  \caption{Lines of constant $\theta$.}
  \label{fig:ne_theta}
\end{figure}

Lines of constant $T$ and constant $\theta$ are shown in
Figs.~\ref{fig:ne_t} and~\ref{fig:ne_theta}.  One observes from
Fig.~\ref{fig:ne_theta} that $\theta$ grows as $N$ decreases, as
expected, and $\theta$ is equal to zero on the periodic instanton line
and on the boundary of the classically allowed region $E_0(N)$.  The
lines of constant $T$ show that $T$ also equals to zero at the
boundary $E_0(N)$, and reaches a maximum for given $N$ (and for given
$\theta$ also) at the bifurcation line $E_1(N)$.  Close to this line
we made use of the modification of the boundary value problem
described at the end of Section~\ref{sec:reg}.

\begin{figure}
  \begin{center}
    \includegraphics{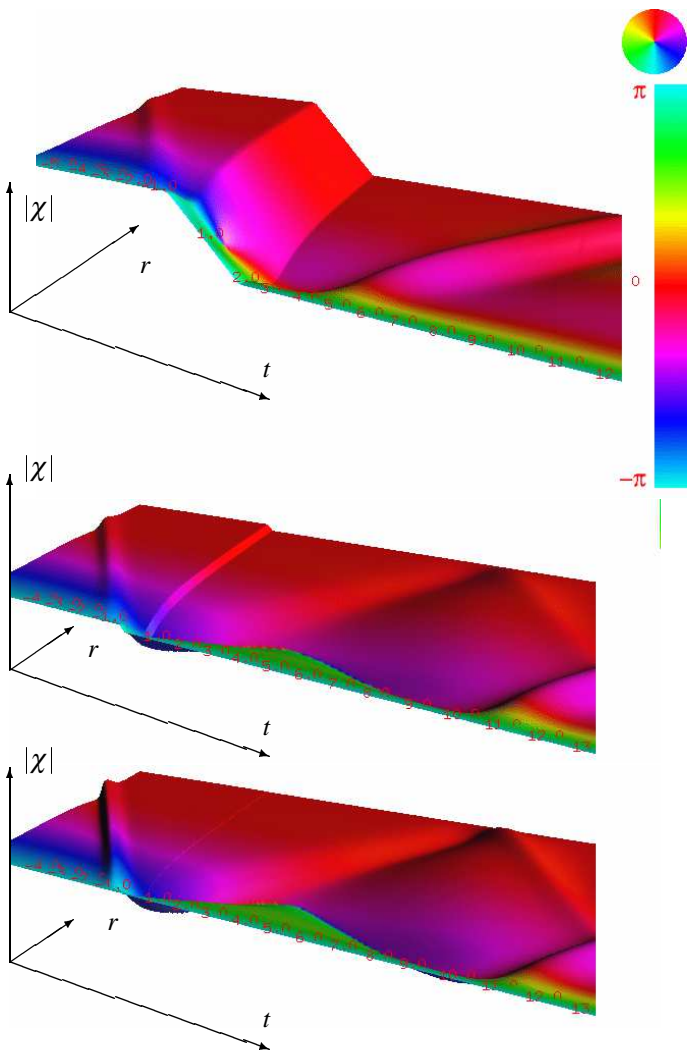}%
  \end{center}
  \caption{Surfaces describing the $\chi$ field for solutions with
    $N=1$ and $E=3.35$ (upper picture), $E=4.48$ (middle), $E=5.22$
    (lower picture).  The first surface corresponds to
    deep underbarrier tunneling, and
    the last one corresponds to nearly classical over-barrier transition.
}
  \label{fig:surfaces}
\end{figure}

\begin{figure}
  \begin{center}
    \includegraphics[width=\columnwidth]{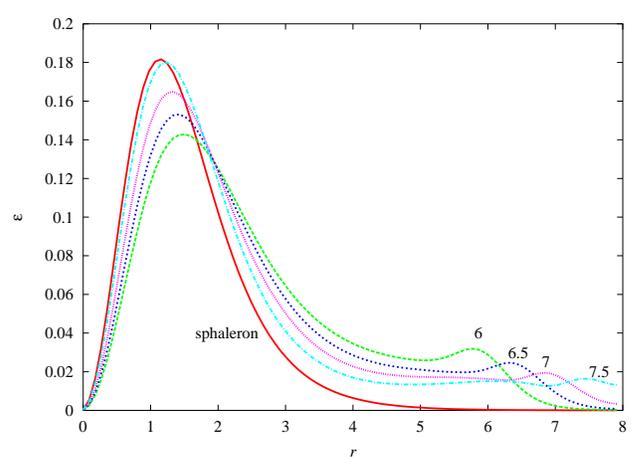}
  \end{center}
  \caption{Energy density for several values of $\Real t$ (indicated
    by numbers near graphs) for the middle configuration of
    Fig.~\ref{fig:surfaces} ($N=1$, $E=3.35$).  The energy density for
    the sphaleron solution is shown for comparison by solid line.}
  \label{fig:sphaleron_decay}
\end{figure}

Representative solutions are shown in Fig.~\ref{fig:surfaces}.  They
correspond to deep tunneling regime ($E<E_1(N)$), tunneling onto the
sphaleron ($E>E_1(N)$) and classical overbarrier transition at
$E_0(N)$, all for $N=1$.  One can see from the color patterns that the
field indeed undergoes the topology changing transition of the form
illustrated in Fig.~\ref{fig:transition}.  The incoming wave is
present in the left part of the pictures, becoming sharper and sharper
for higher energy (the particle number is the same for all plots).
In the first picture the topological transition is seen on the
Euclidean part of the contour.  In the second and third pictures, a
sphaleron like configuration is visible on the right, with ``extra''
waves (excitations about the sphaleron) flying away, while the
sphaleron itself starts to decay quite close to the right end of the
plot (with the regularization parameter $\epsilon$ tending to zero,
the moment of sphaleron decay moves towards larger times).  At large
times, the wave reflected from the boundary $r=L$ appears due to the
Dirichlet boundary conditions~\eqref{lat_space_bc} imposed at $r=L$.
This wave does not alter the results, as it occurs in the
linear regime\footnote{To get rid of the reflected wave a much larger
spatial grid would be needed.}.

The fact that for $E>E_1(N)$ the solution after tunneling has the form
of the sphaleron plus spherical excitations in its background is
illustrated by plotting the spatial energy density at different times
after tunneling.  In Fig.~\ref{fig:sphaleron_decay} the energy density
distribution is shown for the middle solution of
Fig.~\ref{fig:surfaces}.  As the time increases, the bump on the right
(spherical wave) moves towards larger $r$, while the energy density
profile approaches that of the sphaleron.

\begin{figure}
  \centerline{\includegraphics[width=\columnwidth]{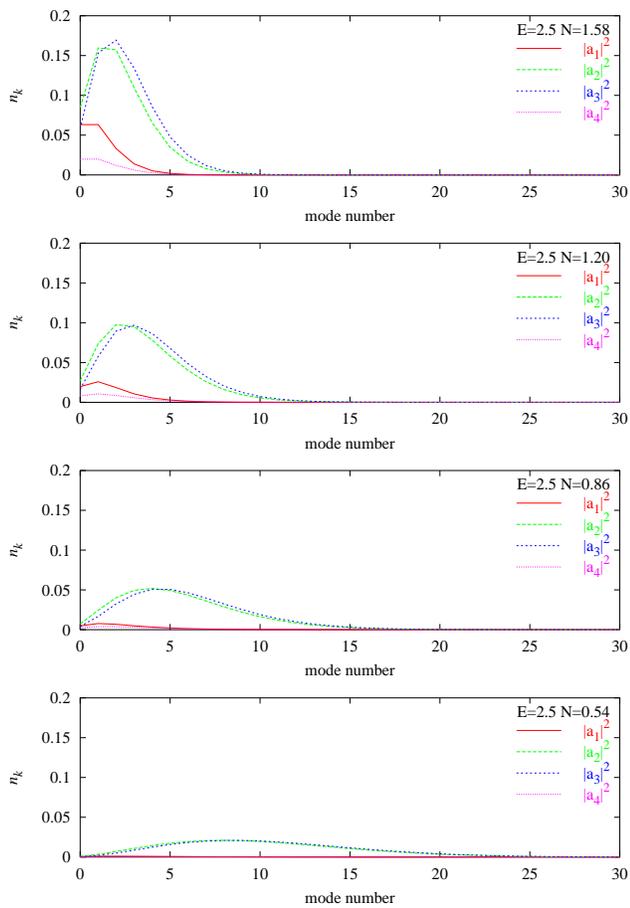}}
  \caption{Distribution of particle number $n_\bk$ at initial time
    over modes for $E\alpha_W/M_W=3.54$ and different $N$.  $a_i$ are
    numbers of particles in each mode in units of $1/\alpha_W$ for
    four different types of modes (see Ref.~\cite{Rebbi:1996zx} for
    definitions).  Mode $a_1$ is the Higgs boson mode, while
    $a_{2,3,4}$ are gauge boson modes ($a_{2,3}$ are transverse, $a_4$
    is radial).  On the horizontal axis is the mode number for a lattice
    with spatial size $r=8$.}
  \label{fig:modes}
\end{figure}

It is also instructive to see that with the number of incoming
particles decreasing, the occupied modes have higher frequencies.
This is demonstrated in Fig.~\ref{fig:modes} for energy slightly
smaller than the sphaleron energy.

Finally, let us discuss the extrapolation of the results to zero
number of particles, which we performed to obtain predictions for
the suppression exponent $F_{HG}(E)$ of the two-particle cross
section.

Two ways of obtaining the lower bounds on $F_{HG}(E)$ were explained
in Section~\ref{sec:formulation}.  One is to continue $F(N)$ at each
energy linearly in $N$ to $N=0$ (this is justified by recalling that
$\partial F/\partial N\propto-\theta$ increases as $N\to0$), while the
other is to continue lines of constant $F$ (see Fig.~\ref{fig:ne_f})
linearly to $N=0$ (this gives a lower bound since the lines of
constant $F$ have positive curvature).  Both these extrapolations are
straightforward to make, insofar as the required derivatives of
$F(E,N)$ are given for each configuration by the values of $T$ and
$\theta$ through the relations~\eqref{dfdn} and~\eqref{dfde}.  In this
way we obtained the bound shown in Fig.~\ref{fig:hg}.

\begin{figure}
  \begin{center}
    \includegraphics[width=\columnwidth]{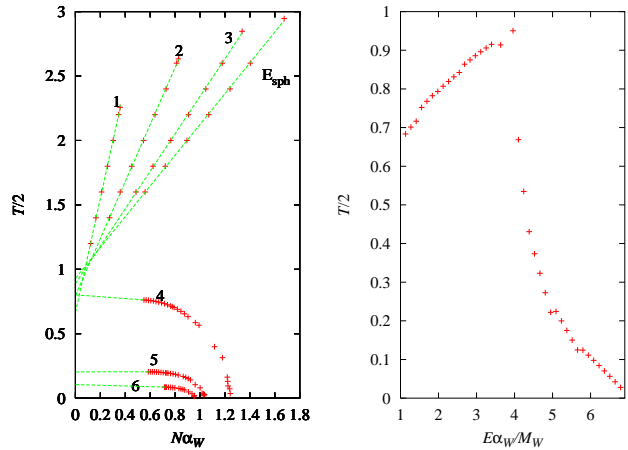}
    \caption{Left: $T(N)/2$ for different energies, labeled by values
      of $E\alpha_W/M_W$, the points are data from numerical
      calculation and lines are extrapolations.  Right: $T(E)/2$
      extrapolated to zero particle number.}
    \label{fig:tn_e}
  \end{center}
\end{figure}

We now elaborate on our estimate of the function $F_{HG}(E)$ itself,
Fig~\ref{fig:hg_low}.  Perturbative calculation at low
energies~\cite{Bezrukov:2003zn} shows that while the exponent $F(E,N)$
has singular behavior of the form $N\log(N)$, the function
$T(E,N)$ is regular in $N$ and close to a linear function.  At large $N$
the numerical data also demonstrate that the behavior of $T$ is close
to linear.  This is shown in Fig.~\ref{fig:tn_e}.  For
$E>E_\mathrm{sph}$ numerical results suggest that $T(N)$ is almost
constant at small $N$.  These properties justify a linear extrapolation
of $T(N)$ to $N=0$ with energy kept constant.  After obtaining $T(E)$
at zero particle number, $F_{HG}$ is readily found by integrating
Eq.~\eqref{dfde} starting from the instanton value $F_{HG}(E=0)=1$.  The
resulting estimate is presented in Fig.~\ref{fig:hg_low}, solid line.

\section{Conclusions}
\label{sec:conclusions}

Our study shows that the semiclassical procedure coupled to suitable
computational techniques is capable of producing quantitative results
for semi-inclusive, weakly coupled non-perturbative processes, and, in
particular, for particle collisions.

In this paper we applied this technique to study the suppression
factor for topology changing transitions, and accompanying baryon and
lepton number violation, in the SU(2) sector of the electroweak theory
up to energies well above the sphaleron energy.
We imposed spatial spherical symmetry, so our results are valid,
strictly speaking, for s-wave scattering.

Our results show that the known analytic expression for the suppression
exponent, which contains three terms of low-energy expansion, works
well up to the sphaleron energy, but underestimates the suppression at
higher energies.

By numerical analysis we have found that baryon and lepton number
violation, accompanying topology changing 
s-wave particle collisions
in the electroweak
theory, remains highly suppressed up to energies of at least $\sim
250$~TeV (and likely much higher).

\begin{acknowledgments}
The authors are indebted to A.~Kuznetsov for helpful discussions.  We
wish to thank Boston University's Center for Computational Science and
Office of Information Technology for generous allocations of
supercomputer time.  Part of this work was completed during visits by
C.R.~at the Institute for Nuclear Research of the Russian Academy of
Sciences and by F.B., D.L., V.R. and P.T. at Boston University, and we
all gratefully acknowledge the warm hospitality extended to us by the
hosting institutions.  The work of PT was supported in part by the
SNSF grant 21-58947.99.  This research was supported by Russian
Foundation for Basic Research grant 02-02-17398, U.S.  Civilian
Research and Development Foundation for Independent States of
FSU~(CRDF) award RP1-2364-MO-02, and DOE grant US DE-FG02-91ER40676.
\end{acknowledgments}


\begin{thebibliography}{10}

\bibitem{Coleman:1977py}
S.~Coleman,
\newblock Phys. Rev. {\bf D15}, 2929 (1977).

\bibitem{Belavin:1975fg}
A.~A. Belavin, A.~M. Polyakov, A.~S. Shvarts and Y.~S. Tyupkin,
\newblock Phys. Lett. {\bf B59}, 85 (1975).

\bibitem{Klinkhamer:1984di}
F.~R. Klinkhamer and N.~S. Manton,
\newblock Phys. Rev. {\bf D30}, 2212 (1984).

\bibitem{Kuzmin:1985mm}
V.~A. Kuzmin, V.~A. Rubakov and M.~E. Shaposhnikov,
\newblock Phys. Lett. {\bf 155B}, 36 (1985).

\bibitem{Arnold:1987mh}
P.~Arnold and L.~McLerran,
\newblock Phys. Rev. {\bf D36}, 581 (1987).

\bibitem{Arnold:1988zg}
P.~Arnold and L.~McLerran,
\newblock Phys. Rev. {\bf D37}, 1020 (1988).

\bibitem{Bochkarev:1987wg}
A.~I. Bochkarev and M.~E. Shaposhnikov,
\newblock Mod. Phys. Lett. {\bf A2}, 991 (1987).

\bibitem{Khlebnikov:1988sr}
S.~Y. Khlebnikov and M.~E. Shaposhnikov,
\newblock Nucl. Phys. {\bf B308}, 885 (1988).

\bibitem{Grigorev:1989je}
D.~Y. Grigoriev, V.~A. Rubakov and M.~E. Shaposhnikov,
\newblock Phys. Lett. {\bf B216}, 172 (1989).

\bibitem{Kuznetsov:1997sf}
A.~N. Kuznetsov and P.~G. Tinyakov,
\newblock Phys. Lett. {\bf B406}, 76 (1997), [hep-ph/9704242].

\bibitem{Frost:1999eh}
K.~L. Frost and L.~G. Yaffe,
\newblock Phys. Rev. {\bf D60}, 105021 (1999), [hep-ph/9905224].

\bibitem{Bonini:2000mb}
G.~F. Bonini {\em et~al.},
\newblock Phys. Lett. {\bf B474}, 113 (2000).

\bibitem{Rubakov:1985am}
V.~A. Rubakov and A.~N. Tavkhelidze,
\newblock Phys. Lett. {\bf B165}, 109 (1985).

\bibitem{Rubakov:1986nk}
V.~A. Rubakov,
\newblock Prog. Theor. Phys. {\bf 75}, 366 (1986).

\bibitem{Matveev:1986bw}
V.~A. Matveev, V.~A. Rubakov, A.~N. Tavkhelidze and V.~F. Tokarev,
\newblock Theor. Math. Phys. {\bf 69}, 961 (1986).

\bibitem{Matveev:1987gq}
V.~A. Matveev, V.~A. Rubakov, A.~N. Tavkhelidze and V.~F. Tokarev,
\newblock Nucl. Phys. {\bf B282}, 700 (1987).

\bibitem{Diakonov:1992mp}
D.~Diakonov and V.~Y. Petrov,
\newblock Phys. Lett. {\bf B275}, 459 (1992).

\bibitem{Rubakov:1985ix}
V.~A. Rubakov,
\newblock JETP Lett. {\bf 41}, 266 (1985).

\bibitem{Ambjorn:1985bb}
J.~Ambjorn and V.~A. Rubakov,
\newblock Nucl. Phys. {\bf B256}, 434 (1985).

\bibitem{Rubakov:1985it}
V.~A. Rubakov, B.~E. Stern and P.~G. Tinyakov,
\newblock Phys. Lett. {\bf 160B}, 292 (1985).

\bibitem{Ringwald:1990ee}
A.~Ringwald,
\newblock Nucl. Phys. {\bf B330}, 1 (1990).

\bibitem{Espinosa:1990qn}
O.~Espinosa,
\newblock Nucl. Phys. {\bf B343}, 310 (1990).

\bibitem{McLerran:1990ab}
L.~McLerran, A.~Vainshtein and M.~Voloshin,
\newblock Phys. Rev. {\bf D42}, 171 (1990).

\bibitem{Khlebnikov:1991ue}
S.~Y. Khlebnikov, V.~A. Rubakov and P.~G. Tinyakov,
\newblock Nucl. Phys. {\bf B350}, 441 (1991).

\bibitem{Yaffe:1990iy}
L.~G. Yaffe,
\newblock Scattering amplitudes in instanton backgrounds,
\newblock in {\em Santa Fe SSC Workshop}, pp. 46--63, 1990.

\bibitem{Arnold:1990va}
P.~B. Arnold and M.~P. Mattis,
\newblock Phys. Rev. {\bf D42}, 1738 (1990).

\bibitem{Mattis:1992bj}
M.~P. Mattis,
\newblock Phys. Rept. {\bf 214}, 159 (1992).

\bibitem{Tinyakov:1993dr}
P.~G. Tinyakov,
\newblock Int. J. Mod. Phys. {\bf A8}, 1823 (1993).

\bibitem{Rubakov:1996vz}
V.~A. Rubakov and M.~E. Shaposhnikov,
\newblock Usp. Fiz. Nauk {\bf 166}, 493 (1996), [hep-ph/9603208].

\bibitem{Rubakov:1992fb}
V.~A. Rubakov and P.~G. Tinyakov,
\newblock Phys. Lett. {\bf B279}, 165 (1992).

\bibitem{Tinyakov:1992fn}
P.~G. Tinyakov,
\newblock Phys. Lett. {\bf B284}, 410 (1992).

\bibitem{Rubakov:1992ec}
V.~A. Rubakov, D.~T. Son and P.~G. Tinyakov,
\newblock Phys. Lett. {\bf B287}, 342 (1992).

\bibitem{Bezrukov:2001dg}
F.~Bezrukov, C.~Rebbi, V.~Rubakov and P.~Tinyakov,
\newblock hep-ph/0110109.

\bibitem{Kuznetsov:1997az}
A.~N. Kuznetsov and P.~G. Tinyakov,
\newblock Phys. Rev. {\bf D56}, 1156 (1997), [hep-ph/9703256].

\bibitem{Mueller:1993sc}
A.~H. Mueller,
\newblock Nucl. Phys. {\bf B401}, 93 (1993).

\bibitem{Bonini:1999cn}
G.~F. Bonini, A.~G. Cohen, C.~Rebbi and V.~A. Rubakov,
\newblock quant-ph/9901062.

\bibitem{Bonini:1999kj}
G.~F. Bonini, A.~G. Cohen, C.~Rebbi and V.~A. Rubakov,
\newblock Phys. Rev. {\bf D60}, 076004 (1999), [hep-ph/9901226].

\bibitem{Bezrukov:2003yf}
F.~Bezrukov and D.~Levkov,
\newblock quant-ph/0301022.

\bibitem{'tHooft:1976fv}
G.~'t~Hooft,
\newblock Phys. Rev. {\bf D14}, 3432 (1976).

\bibitem{Espinosa:1992vq}
O.~R. Espinosa,
\newblock Nucl. Phys. {\bf B375}, 263 (1992).

\bibitem{Rebbi:1996zx}
C.~Rebbi and J.~Singleton, Robert,
\newblock Phys. Rev. {\bf D54}, 1020 (1996), [hep-ph/9601260].

\bibitem{Akiba:1989xu}
T.~Akiba, H.~Kikuchi and T.~Yanagida,
\newblock Phys. Rev. {\bf D40}, 588 (1989).

\bibitem{Khlebnikov:1991th}
S.~Y. Khlebnikov, V.~A. Rubakov and P.~G. Tinyakov,
\newblock Nucl. Phys. {\bf B367}, 334 (1991).

\bibitem{Voloshin:1994dk}
M.~B. Voloshin,
\newblock Phys. Rev. {\bf D49}, 2014 (1994).

\bibitem{Bezrukov:2003zn}
F.~Bezrukov and D.~Levkov,
\newblock hep-th/0303136.

\bibitem{Ringwald:2002sw}
A.~Ringwald,
\newblock Phys. Lett. {\bf B555}, 227 (2003), [hep-ph/0212099].

\bibitem{Ringwald:2003px}
A.~Ringwald,
\newblock hep-ph/0302112.

\bibitem{Khoze:1991bm}
V.~V. Khoze and A.~Ringwald,
\newblock Nucl. Phys. {\bf B355}, 351 (1991).

\bibitem{Arnold:1991cx}
P.~B. Arnold and M.~P. Mattis,
\newblock Mod. Phys. Lett. {\bf A6}, 2059 (1991).

\bibitem{Diakonov:LINPSchool1991}
D.~I. Diakonov and V.~Y. Petrov,
\newblock in {\em Proc.\ XXVI LINP Winter School. LINP, Leningrad}, 1991.

\bibitem{Mueller:1991fa}
A.~H. Mueller,
\newblock Nucl. Phys. {\bf B364}, 109 (1991).

\bibitem{Ringwald:1999ze}
A.~Ringwald and F.~Schrempp,
\newblock Phys. Lett. {\bf B459}, 249 (1999), [hep-lat/9903039].

\bibitem{Schrempp:2002kd}
F.~Schrempp and A.~Utermann,
\newblock Phys. Lett. {\bf B543}, 197 (2002), [hep-ph/0207300].

\bibitem{Ratra:1988dp}
B.~Ratra and L.~G. Yaffe,
\newblock Phys. Lett. {\bf B205}, 57 (1988).

\bibitem{Farhi:1996aq}
E.~Farhi, J.~Goldstone, A.~Lue and K.~Rajagopal,
\newblock Phys. Rev. {\bf D54}, 5336 (1996), [hep-ph/9511219].

\end{thebibliography}

\end{document}